\documentclass[sn-mathphys]{sn-jnl}

\begin{document}

\title[Article Title]{Observational constraints on Tsallis holographic dark energy with Ricci horizon cutoff}

\author*[1,2]{\fnm{Zahra} \sur{Feizi Mangoudehi}}\email{zahraf@msc.guilan.ac.ir}

\affil*[1]{\orgdiv{Department of Physics}, \orgname{University of Guilan}, \orgaddress{\city{Rasht}, \country{Iran}}}

\abstract{In this research, we are interested in constraining the nonlinear interacting and noninteracting Tsallis holographic dark energy (THDE) with Ricci horizon cutoff by employing three observational datasets.
To this aim,  the THDE with Ricci horizon considering the noninteraction and nonlinear interaction terms will be fitted by the SNe Ia, SNe Ia+H(z), and SNe Ia+H(z)+GRB samples to investigate the Hubble ($H(z)$), dark energy equation of state ($\omega_{DE}$), effective equation of state ($\omega_{eff}$), and deceleration ($q$) parameters.
Investigating the $H(z)$ parameter illustrates that our models are in good consistent with respect to observation. Also, it can reveal the turning point for both noninteracting and nonlinear interacting THDE with Ricci cutoff in the late time era.

Next, the analysis of the $\omega_{DE}$ for our models displays that the dark energy can experience the phantom state at the current time. However, it lies in the quintessence regime in the early era and approaches the cosmological constant in the late-time epoch. Similar results will be for the $\omega_{eff}$ parameter with this difference that the $\omega_{eff}$ will experience the quintessence region at the current redshift.

Next to the mentioned parameters, the study of the $q$ parameter indicates that the models satisfy an acceptable transition phase from the matter to the dark energy-dominated era.

After that, the classical stability ($v_{s}^{2}$) will be analyzed for our models. The $v_{s}^{2}$ shows that the noninteracting and nonlinear interacting THDE with Ricci cutoff will be stable in the past era, unstable in the present, and progressive epochs.
Then, we will employ the $Jerk$ ($J$) and $OM$ parameters to distinguish between our models and the $\Lambda CDM$ model.
Finally, we will calculate the age of the Universe for the THDE and nonlinear interacting THDE with Ricci as the IR cutoff.
 }


\maketitle

\section{Introduction}\label{sec1}

Today, analysis obtained from SNe Ia (\cite{bib1,bib2,bib3,bib4}), CMB (\cite{bib5,bib6,bib7,bib8,bib9}), and BAO
(\cite{bib10}) data confirms that the Universe is expanding with acceleration features. This characteristic of the
Universe can be described by introducing a source of energy with repulsive gravitational behavior (caused by negative
pressure) called dark energy (DE). Although the nature of dark energy is not evident, some candidates, such as the
cosmological constant $\Lambda$, are counted as the most favorable dark energy with a famous equation of state (Eos)
parameter -1. However, it suffers coincidence and fine-tuning (\cite{bib11,bib12,bib13,bib14,bib15}) problems. Therefore,
Many alternative dark energy candidates have been introduced to approach these difficulties.

So, one of the most significant dark energy models suggested by \cite*{bib16} is called the holographic dark energy
(HDE) model. This model is based on the \cite*{bib17} suggestion representing that in the effective quantum field theory, the short-distance or UV cutoff ($\Lambda$) is related to the long-distance or IR cutoff ($L$) resulting from the limit set by the black hole information. Thus, the quantum zero-point energy ($\rho_{\Lambda}$) of a system with size $L$ should not be upper than the mass of a black hole of the same size as

\begin{eqnarray}\nonumber
L^{3}\rho_{\Lambda}=M_{pl}^{2}L.
\end{eqnarray}

From this relation, \cite*{bib32} introduced the following holographic dark energy density:

\begin{eqnarray}\nonumber
\rho_{HDE}=3c^{2}M_{pl}^{2}L^{-2}.
\end{eqnarray}

Where $3c^{2}$ is a numerical constant and $L$ is the size (or horizon) of the current Universe. \cite*{bib16} has examined the HDE model with particle and future event (FE) horizons. The author found that the $\omega_{DE}$ for $c=1$ and $c<1$ with the FE cutoff will be -1 and
$<-1$.

Later, \cite*{bib18} constrained the $\Omega_{m}$, $\omega_{HDE}$, and $z_{tr}$ (the transition redshift) parameters
using the supernova type Ia data for fixed $c=1$ and free $c$ for the HDE with FE cutoff. The fitted parameters have been measured as $\Omega_{m}=0.25$, $\omega_{HDE}=-0.91$, and $z_{tr}=0.72$  for the model with fixed $c=1$ and $\Omega_{m}=0.46$, $\omega_{HDE}=-2.67$, $c=0.21$, and $z_{t}=0.28$ for the model with non-fixed $c$.

Apart from holographic dark energy, in 2007,
Cai introduced the agegraphic dark energy (ADE) by applying the quantum fluctuations of spacetime
and the K\'{a}rolyh\'{a}zy relation. In this model, the author has applied the age of the Universe as the infrared cutoff
led to the ADE density described by (see Ref. \cite*{bib19})

\begin{eqnarray}\nonumber
\rho_{q}=3n^{2}M_{pl}^{2}T^{-2}.
\end{eqnarray}

Then, the $\omega_{DE}$ parameter for the ADE model in the Universe filled by dark energy component has been derived.  The author found that if  $n>1$, the Universe will be in the accelerated expansion phase. Moreover, he measured the present
value of $\omega_{DE}$ ($\omega_{DE0}$) for the Universe consisting of matter and ADE components with the result of
$\omega_{DE0}\leq-0.81$ for $n\geq3$.

Then, \cite*{bib20} using the metric fluctuations, the K\'{a}rolyh\'{a}zy relation, and the conformal time ($\eta$) proposed a new agegraphic dark energy (NADE) with $\rho_{q}=3n^{2}M_{pl}^{2}\eta^{-2}$. Authors have investigated the
$\omega_{DE}$ and $\Omega_{DE}$ parameters in the NADE with the non-coupling term. Then, at the continuation of their work, they have calculated the $\Omega_{DE}$ parameter with coupling ingredient between dark sectors. The noninteracting model implies $\omega_{DE}=-\frac{2}{3}$ and $\Omega_{DE}=\frac{n^{2}a^{2}}{4}$ in the matter-dominated epoch and $\omega_{DE}=-\frac{1}{3}$ and $\Omega_{DE}=n^{2}a^{2}$ in the radiation era. So, the $\omega_{DE}$ in this non-coupling model cannot enter the phantom state. Despite it, the NADE with the interacting part leads the DE to pass the phantom divide line.

After this work, \cite*{bib21} has studied the NADE in the framework of f(R) gravity. In the first, he has derived the
equation of state parameter for the model and observed that the $\omega_{DE}$ can pass the -1 for $n < 0$. And in the last, he has reconstructed the $f(R)$ function using the general $f(R)$ gravity action and new agegraphic dark energy density.

A few years later, \cite*{bib22} introduced a new holographic dark energy deriving the non-additive entropy for complex
systems, including black holes (\cite*{bib23}). They have employed the Tsallis horizon entropy as

\begin{eqnarray}
S_{\delta}=\gamma A^{\delta},
\end{eqnarray}

where $\delta$ and $\gamma$ are non-additivity and unknown constant parameters, respectively. Applying the holographic
principle, the relation between IR and UV cut-offs, and Eq. (1) gives

\begin{eqnarray}\nonumber
\Lambda^{4}\leq (\gamma(4\pi)^{\delta})L^{2\delta-4}.
\end{eqnarray}

Here, $\Lambda^{4}$ is the vacuum energy density that creates the acceleration of the Universe.

So, the new holographic dark energy density is given by

\begin{eqnarray}
\rho_{DE}=B L^{2\delta-4},
\end{eqnarray}

called Tsallis holographic dark energy (THDE). The authors have applied the THDE with Hubble horizon supposing
$\delta$= 1.4, 1.5, and 1.7 with $\Omega_{DE0}$= 0.70 and $\Omega_{DE0}$= 0.73 to measure the $\omega_{DE}$,
$\Omega_{DE}$, $q$, and $\omega_{eff}$. The results show that at $z\longrightarrow\infty$ ($z\longrightarrow-1$)
$\Omega_{DE}\longrightarrow0$ ($\Omega_{DE}\longrightarrow1$), and $\omega_{DE}\longrightarrow1-\delta$
($\omega_{DE}\longrightarrow-1$). Also, for $\delta=2$, the $\omega_{DE}$ is -1. From studying the $q$ parameter,
$z\longrightarrow\infty$, $q\longrightarrow0.5$, while at $z\longrightarrow-1$, $q$ tends to -1. Additionally, the
model has been studied in terms of $v_{s}^{2}$ concluding that the model is unsustainable for $\delta>1$ and stable for
$\delta\leq1$. Moreover, the age of the current Universe has been estimated for the model. For
instance, $t\approx\frac{1.5}{H_{0}}$ when $\delta=1.7$, $\Omega_{DE0}=0.70$, and $\omega_{DE}\simeq-0.87$. This work
with the Hubble cutoff satisfies the accelerating state of the Universe.

After this work, \cite*{bib24} studied the THDE with various IR cutoffs, including the particle, Ricci, and
Granda-Oliveros (GO) horizons for the noninteracting and (linear) interacting cases. Although the Hubble cutoff has
been chosen for only interacting THDE model. The authors have studied the evolution of the $\Omega_{DE}$, $\omega_{DE}$,
$q$, and $v_{s}^{2}$ in models with $\Omega_{DE0}=0.73$ and $H_{0}=67$. Here, we will present some results, such
as $\omega_{DE}$ and $v_{s}^{2}$ features. The evolution of the $\omega_{DE}$ states that the model with Hubble cutoff
for $b^{2}$= 0.03, 0.04, 0.05, and $\delta=1.4$ and in addition, the Ricci cutoff without $Q$ (with $Q$) for $\lambda=1$
and $\delta=1$ ($\lambda$= 1 with $b^{2}=0.01$ and $\delta=1$) can display passing from the quintessence to the
phantom regime.

Then, the authors found the positive sign of  $v_{s}^{2}$ at $z\longrightarrow\infty$ for the following models: First, the model with particle cutoff in the presence of coupling term $Q$ for different $\delta$ values, $b^{2}=0.1$, $B=2.4$, or various $b^{2}$ with $\delta=2.4$, and $B=2.4$. Moreover, the noninteracting and interacting models with GO cutoff taking $\delta=0.6$, $\alpha=0.8$, and $\beta=0.5$ and $b^{2}=0$ or $b^{2}=0.01$. And then, the THDE model with the Ricci cutoff with $Q=0$ for $\lambda=1$ and $\delta=1$.

Later, \cite*{bib25} constrained the THDE with Hubble and future event horizons using the Pantheon SN Ia, BAO, CMB,
GRB, and a local sample of H, for noninteracting and linear interacting cases (with $Q=3Hb\rho_{m}$). The parameters which have been fitted are $H_{0}$, $\Omega_{DE}$, $\delta$, $b\neq0$ or $b=0$, $z_{tr}$, and Age (of the present Universe). Using the fitted values, the author has surveyed the Information Criterion (IC), Alcock-Paczynski (AP) test, and stability situation in the THDE model. The results of IC show that selecting HDE as the reference model led to supporting the THDE and ITHDE by observational data. From the AP test, the THDE and ITHDE with Hubble (future event) horizon separate from each other at $z$=1.2 (1.5) and $z$=0.7 (1.05). Moreover, the THDE with FE cutoff shows the lowest deviation versus the $\Lambda CDM$ and HDE rather than THDE with Hubble cutoff. Investigating stability represents that although models with Hubble cutoff in all eras are unsustainable (similar to \cite*{bib1}), models with FE cutoff at $z\longrightarrow-1$ will show sustainability.

\cite*{bib26} has discussed the THDE model with GO cutoff in ($n$+1) dimensional FLRW Universe. The model is analyzed
via cosmographic parameters, the statefinder pair, the $Om$ diagnostic, and the diagnostic parameter $L_{m}^{1}$ with
$\alpha$=0.8502 and $\beta$=0.4817. The $Om$ plot shows the model is in a positive slope relating to the phantom era and the $L_{m}^{1}$ graph with  $L_{m}^{1}\neq0$ illustrates the model is not a $\Lambda CDM$ model. In addition, the author has found the $\omega_{DE}$ and $V_{s}^2$ of the model for noninteracting and interacting cases. The $\omega_{DE0}$ is observed in the phantom regime in different redshifts, and $v_{s}^{2}$ has demonstrated a sustainable behavior with initial conditions $\Omega_{DE}=0.70$, $H_{0}=73$, $\alpha=0.8502$,
and $\beta=0.4817$.

As well as this, \cite*{bib27} studied the correspondence between the THDE and Generalized Chaplygin
Gas (GCG) in the framework of Compact Kaluza-Klein cosmology. On the ground, considering $-0.05\leq b^{2}\leq-0.03$ and $\delta=2.2$, the $\Omega_{DE}$, $\omega_{DE(eff)}$, $q$, and $v_{s}^{2}$ have been plotted. Moreover, the plane of $\omega_{DE(eff)}-\Omega_{DE}$ with $b^{2}>0$ and $b^{2}<0$ (with $b^{2}$= $\pm$0.20, $\pm$0.25, $\pm$0.28) and fixed value of $\delta$ including 0.2, 0.3, 0.4, or 0.5 have been depicted. Also, the $v_{s}^{2}-\Omega_{DE}$  plot has been graphed for different $\delta$ (or $0.3\leq\delta\leq0.5$), once by fixing $b^{2}$= -0.20 and then $b^{2}$= -0.25. Evolving the dark energy EoS parameter versus redshift expresses that for  $b^{2}<0$ ($b^{2}>0$), the $\omega_{DE(eff)}$ has started from the phantom-like ($\omega_{DE(eff)}<-1$) phase in the past to the quintessence (phantom) region at $\Omega_{DE}\longrightarrow0.80$. For the $\omega_{DE}$ with $b^{2}<0$ and $\delta=2.2$, the dark energy behaves as the phantom at $z\longrightarrow \infty$ and $z\longrightarrow 0$, then approaches -1 at $z\longrightarrow -1$. From the $v_{s}^{2}$ plots, the model is unstable for $\delta= 2.2$ and $b^{2}<0$, but it will be stable with $0.3\leq\delta\leq0.5$ and $b^{2}<0$. For $\delta>0.63$, independent of the value of $b^{2}$, the sign of  $v_{s}^{2}$ is negative.

Subsequently, \cite*{bib28} has regarded the THDE model in the Brans-Dicke scenario. To solve the field equations and
the wave equation of the scalar field ($\phi$), the author has selected the scalar field as a function of the scale factor and a constant as $\phi$=$a^{n}$. Then, solving the wave equation gives $n= \frac{3}{2\omega}$, $n^{2}+3n=\frac{3}{\omega}$, and $k=0$. By choosing $\omega= -\frac{3}{2}$ and $n=-1$, the scalar field leads to consistent results. However, the relation between the scalar field and density causes the conservation law for energy can not be held.

In another research work, \cite*{bib29} have worked on the THDE with Ricci horizon cutoff. The structure of this work
is different from the \cite*{bib24} project. Here, the authors have taken ansatz for the scale factor ($a=t^{k}e^{t}$) to obtain the $H(t)$ (Hubble parameter), $\omega_{DE}$, $\rho_{m}$ (matter density), $\rho_{DE}$ (dark energy density), $P_{DE}$ (dark energy pressure), $\Omega_{DE}$ (DE density parameter), and $v_{s}^{2}$. The plot of $\omega_{DE}$ with $\lambda=1.5$ and $k=0.6$ from $z\longrightarrow\infty$ to $z\longrightarrow0$ indicates that the DE behaves as the quintessence (phantom) for $\delta$= 1.8 and 1.9 ($\delta$= 2.1 and 2.2). However, they will tend to be -1 in the future.
As well as this, supposing $\lambda=1.5$ and $k=0.6$ as a fixed values, the sign of  $v_{s}^{2}$ for $\delta$=1.8, 1.9 (or $\delta$=2.1, 2.2) is $>0$ ($<0$) at $z\longrightarrow\infty$, while it is $<0$ ($>0$) at $z\longrightarrow0$ and $z\longrightarrow-1$.The results of $v_{s}^{2}$ for $\lambda$= 1.0, 1.5, and 2.0 as the free values show the $v_{s}^{2}>0$ ($v_{s}^{2}<0$) for $\delta$= 1.8, 1.9 ($\delta$= 2.1, 2.2) at $z\longrightarrow\infty$ and $v_{s}^{2}<0$ ($v_{s}^{2}>0$) at $z\longrightarrow0$ and $z\longrightarrow-1$.
Then, the authors have examined the THDE model with the quintessence, phantom, and k-essence fields. These fields can describe the expansion of the Universe with acceleration behavior.

So, motivated by these studies, we are interested in exploring the observational constraints on the Tsallis
holographic dark energy regarding the Ricci scalar as IR cutoff without interaction and with nonlinear interaction
components using three observational samples. For this purpose, we will fit the
parameters of the models applying SNe Ia, SNe Ia+H(z), and SNe Ia+H(z)+GRB data. The confidence counters and likelihood
plots (1-dimension) for the fitted models will be graphed.
Then, the models will be tested through the evolution of the Hubble plane against the observational data and searching for the turning point feature. Additionally, we study the evolution of $\omega_{DE}$, $\omega_{eff}$, and $q$ parameters. Following them, the $v_{s}^{2}$ will be studied. Next, $Jerk$ and $OM$ diagnostic parameters will be investigated. Finally, the age of the Universe will be calculated.

This manuscript is organized as follows: Sec. 2 represents the structure of the THDE model without interaction and with a nonlinear interaction term between dark matter and dark energy by considering the Ricci cutoff. In sec. 3, the best-fitted values of the models for SNe Ia, SNe Ia+H(z), and SNe Ia+H(z)+GRB samples will be measured. In addition, we will examine the parameters' accuracy of the models with confidence graphs and 1-dim likelihood plots. In section 4, the Hubble graph will be studied. In section 5, the dark energy equation of state, effective equation of state, and deceleration parameters will be investigated. Section 6 is devoted to studying the classical stability, and section 7 to the $Jerk$ and $OM$ diagnostic parameters. In sec. 8, the age of the present Universe will be estimated. Finally, the conclusion is prepared in the last section.

\section{The models}\label{sec2}

The Universe with a flat homogenous and isotropic characteristic can be described by the
Friedmann-Lema\^{i}tre-Roberston-Walker (FLRW) metric,

\begin{eqnarray}\nonumber
ds^{2}=-dt^{2}+a^2(t)(dr^{2}+r^{2}d\Omega^{2}),
\end{eqnarray}

in which, $a(t)$ is the scale factor. So, The Universe obeys the following Friedmann equation:

\begin{eqnarray}
3H^{2}=\frac{1}{M_{pl}^{2}}(\rho_{DE}+\rho_{m}).
\end{eqnarray}

Here, $\rho_{m}$ and $\rho_{DE}$ are the matter and dark energy densities. These densities follow the conservative
equation shown by

\begin{eqnarray}
&&\dot{\rho}_{m}+3H(\rho_{m}+p_{m})=Q,  \\
&&\dot{\rho}_{DE}+3H(\rho_{DE}+p_{DE})=-Q.
\end{eqnarray}

Where the matter is pressureless (or $p_{m}$=0) and dark energy is with $P_{DE}=\omega_{DE}\rho_{DE}$. Here, $Q$ is
the interaction term between dark sectors. This term has been suggested by recent observational data (as an
example \cite{bib30,bib31}) and can alleviate the coincidence problem (\cite{bib32,bib11,bib33,bib34,bib35}).
Therefore, many researchers have studied different types of interacting models, while in some cases, noninteracting
and interacting models are compared (\cite{bib35,bib36,bib37,bib38,bib32,bib39,bib40,bib24,bib25,bib41,bib42,bib43,bib44}).

In the previous works of the THDE with Ricci horizon cutoff, \cite*{bib24} have focused on the model with the
noninteraction ($Q=0$) and linear interaction ($Q=3b^{2}H(\rho_{m}+\rho_{DE})$) terms. However, \cite*{bib29} have
worked on the model with $Q=0$. In the present research, we would like to explore the effects of the nonlinear
interaction term $Q= 3H\psi\frac{\rho_{m}\rho_{DE}}{\rho_{m}+\rho_{DE}}$ (\cite{bib39,bib40}) on the THDE with Ricci cutoff. Although, we will focus on the model with $Q=0$, as well.

The matter and DE densities with respect to $\rho_{cr}$ (the critical energy density) will give us

\begin{eqnarray}
\Omega_{m}=\frac{\rho_{m}}{3M_{pl}^{2}H^{2}} ,~\ \ \Omega_{DE}=\frac{\rho_{DE}}{3M_{pl}^{2}H^{2}},
\end{eqnarray}

which are the matter and dark energy density parameters.

Now, we write the Ricci horizon cutoff (\cite{bib45})

\begin{eqnarray}
L=(\alpha H^{2}+\beta \dot{H})^\frac{-1}{2},
\end{eqnarray}

in which, $\alpha=2\beta$. So, from relations (2) and (7), the THDE density can be written (\cite{bib24,bib46})

\begin{eqnarray}
\rho_{DE}=\lambda (2 H^{2}+\dot{H})^{(2-\delta)}.
\end{eqnarray}

Where, $\lambda= B \beta^{2-\delta}$. Derivative from the Eqs. (3) and (6) with respect to $X=\ln{a}$ and inserting Eq. (4), and $\Omega_{m}=1-\Omega_{DE}$ for the noninteracting model we have

\begin{eqnarray}
\frac{d\Omega_{DE}}{dX}=(1-\Omega_{DE})(2\frac{1}{H}\frac{dH}{dX}+3),
\end{eqnarray}\\

and for nonlinear model,

\begin{eqnarray}
\frac{d\Omega_{DE}}{dX}=(1-\Omega_{DE})(2\frac{1}{H}\frac{dH}{dX}+3-3\psi\Omega_{DE}).
\end{eqnarray}\\

Where (\cite{bib24,bib46})

\begin{eqnarray}
\frac{dH}{dX}=H(\frac{(3 H^{2}\Omega_{DE})^\frac{1}{2-\delta}}{H^{2} \lambda^\frac{1}{2-\delta}}-2),
\end{eqnarray}

In the following section, we apply the obtained equations.

\section{Constraints of the model parameters by observational data}\label{sec3}

In this section, we constrain the set of free parameters of the model ($\theta$) with $\theta$=($h$, $\Omega_{DE}$,
$\delta$, $\lambda$, $\psi$) for the nonlinear interacting THDE model and $\theta$=($h$, $\Omega_{DE}$, $\delta$,
$\lambda$) for the noninteracting THDE model by observational data.

Here, we employ the 1048 Supernovae type Ia data (\cite{bib47}) called the Pantheon sample. Then, the Pantheon sample is combined with the 27 Hubble parameter measurements (see Table 1) as SNe Ia + H(z). Finally, the 162 GRB data (\cite{bib48}) will be added to the Pantheon and H(z) data points (or SNe Ia + H(z) + GRB ).

Next, to fit the models with the observational samples, we will minimize the $\chi^{2}$ function which is

\begin{eqnarray}\nonumber
\chi(\theta)_{(SNe Ia)}^{2}=\sum_{i=1}^{1048}[\frac{\mu(z_{i})_{obs}-\mu(z_{i},
\theta)_{th}}{\sigma_{\mu(z_{i})_{obs}}}]^{2},
\end{eqnarray}

\begin{eqnarray}\nonumber
\chi(\theta)_{(H(z))}^{2}=\sum_{i=1}^{27}[\frac{H(z_{i},\theta)_{th}-H(z_{i})_{obs}}{\sigma_{H(z_{i})_{obs}}}]^{2},
\end{eqnarray}

and

\begin{eqnarray}\nonumber
\chi(\theta)_{(GRB)}^{2}=\sum_{i=1}^{162}[\frac{\mu(z_{i})_{obs}-\mu(z_{i},
\theta)_{th}}{\sigma_{\mu(z_{i})_{obs}}}]^{2}.
\end{eqnarray}

Where $\mu(z)_{obs}$ and $\mu(z)_{th}$ are the observed distance modulus and the theoretical distance
modulus.

Also,

\begin{eqnarray}\nonumber
&&\mu(z)_{th}=5 log_{10}d_{L}(z)+\mu_{0} ,~\ \  \mu_{0}=42.38-5 log_{10}h ,~\ \
\nonumber\\&&d_{L}(z)=(1+z)\int_{0}^{z}(\frac{dz}{H(z)})\nonumber
\end{eqnarray}

The total $\chi(\theta)_{min}^{2}$ for the datasets will be $\chi(\theta)_{(SNe Ia)}^{2}$, $\chi(\theta)_{(SNe
Ia)}^{2}$+$\chi(\theta)_{(H(z))}^{2}$, and $\chi(\theta)_{(SNe Ia)}^{2}$+$\chi(\theta)_{(H(z))}^{2}$+$\chi(\theta)_{(GRB)}^{2}$.\\

Then, the likelihood function can be calculated with

\begin{eqnarray}\nonumber
L_{max}(\theta)=e^{-\frac{(\chi(\theta)_{min}^{2})}{2}}
\end{eqnarray}

Our fitting process is done by an algorithm including for...do loops in maple software. We use for...do loop for every free parameter of the models. For example, in the noninteracting THDE with Ricci cutoff with four free parameters ($h$, $\Omega_{DE}$, $\delta$,
$\lambda$), we will use four for...do loops. In more details, for $\delta$ we have: for delta from 0.12 by 0.01 to 0.14 do. Other for...do loops are applied for $\chi^{2}$ function (e.g., eight for...do loops for SNe Ia data). To find the best fit values, first, we need to choose intervals for parameters of the models to put inside the for...do loops (such as $0.1\leqslant\delta\leqslant1.3$ for $\delta$ parameter). By changing the values of intervals, the value of $\chi^{2}_{min}$ will increase or decrease.

Finding the best-fit parameters need time and hardworking attempts. But, after reaching them, we have to examine the results again. For instance, if our best fit values and the minimum value of $\chi^{2}$ extracted from $0.702\leqslant h\leqslant0.703$ (with 0.001 step), $0.71\leqslant \Omega_{DE}\leqslant0.72$ (with 0.0001 step), $0.12\leqslant \delta\leqslant0.14$ (with 0.01 step), and $0.0004\leqslant \lambda\leqslant0.0006$ (with 0.0001 step), with every changes in steps we will understand our results are true or not. If the results of fitting models show that our free values will remain inside intervals and $\chi^{2}_{min}$ does not increase or decrease, then we could find a suitable and reliable best-fit result.

\begin{table}[t]
\centering
\caption{27 H(z) data points respect to redshift $z$. References are: 1) \cite*{bib49}, 2) \cite*{bib50}, 3)
\cite*{bib51}, 4) \cite*{bib52}, 5) \cite*{bib53}, 6) \cite*{bib54}}
\begin{tabular}{|@{}llll|}
\hline
~~$z$ ~~~&~~~~~$H$~~~~~~&$\sigma_{H}$ ~~&~~Reference\\ [2ex] 
\hline 
~~0.070 ~~~~&~~~~ 69.0~~& 19.6 ~~~&~~~~~ 1\\ 
\hline 
~~0.090 ~~~~&~~~~ 69& 12.0 ~~~&~~~~~ 2 \\ 
\hline 
~~0.120 ~~~~&~~~~ 68.6 ~~& 26.2 ~~~&~~~~~~1\\ 
\hline 
~~0.170 ~~~~&~~~~ 83~~& 8 ~~~&~~~~~ 2 \\ 
\hline 
~~0.179 ~~~~&~~~~ 75 ~~& 4 ~~~&~~~~~ 3\\ 
\hline 
~~0.199 ~~~~&~~~~ 75~~& 5 ~~~&~~~~~ 3\\ 
\hline 
~~0.200 ~~~~&~~~~ 72.9 ~~& 29.6 ~~~&~~~~~ 1\\ 
\hline 
~~0.270 ~~~~&~~~~ 77~~& 14 ~~~&~~~~~ 2\\ 
\hline 
~~0.280 ~~~~&~~~~ 88.8 ~~& 36.6 ~~~&~~~~~ 1\\ 
\hline 
~~0.352 ~~~~&~~~~ 83 ~~& 14 ~~~&~~~~~ 3\\ 
\hline 
~~0.3802~~~~&~~~~ 83 ~~& 13.5 ~~~&~~~~~ 4\\ 
\hline 
~~0.400 ~~~~&~~~~ 95~~& 17 ~~~&~~~~~ 2\\ 
\hline 
~~0.4004 ~~~~&~~~~ 77~~& 10.2 ~~~&~~~~~~4 \\ 
\hline 
~~0.4247 ~~~~&~~~~ 87.1~~& 11.2 ~~~&~~~~~~4 \\ 
\hline 
~~0.4497~~~~&~~~~ 92.8~~& 12.9 ~~~&~~~~~ 4\\ 
\hline 
~~0.4783 ~~~~&~~~~ 80.9  ~~& 9 ~~~&~~~~~ 4\\ 
\hline 
~~0.480 ~~~~&~~~~ 97 ~~& 62 ~~~&~~~~~ 5\\ 
\hline 
~~0.593  ~~~~&~~~~ 104~~& 13 ~~~&~~~~~ 3\\ 
\hline 
~~0.680 ~~~~&~~~~ 92 ~~& 8 ~~~&~~~~~ 3\\ 
\hline 
~~0.781 ~~~~&~~~~ 105 ~~& 12 ~~~&~~~~~ 3\\ 
\hline 
~~0.875 ~~~~&~~~~ 125 ~~& 17 ~~~&~~~~~ 3\\ 
\hline 
~~0.880 ~~~~&~~~~ 90 ~~& 40 ~~~&~~~~~ 5\\ 
\hline 
~~0.900 ~~~~&~~~~ 117 ~~& 23 ~~~&~~~~~ 2\\ 
\hline 
~~1.037 ~~~~&~~~~ 154 ~~& 20 ~~~&~~~~~ 3\\ 
\hline 
~~1.300 ~~~~&~~~~ 168 ~~& 17 ~~~&~~~~~ 2\\ 
\hline 
~~1.363 ~~~~&~~~~ 160 ~~& 33.6 ~~~&~~~~~ 6\\ 
\hline 
~~1.43 ~~~~&~~~~ 177 ~~& 18 ~~~&~~~~~ 2\\ 
\hline
\end{tabular}
\end{table}

\begin{figure*}
\includegraphics[width=2.1cm, height=2.1cm]{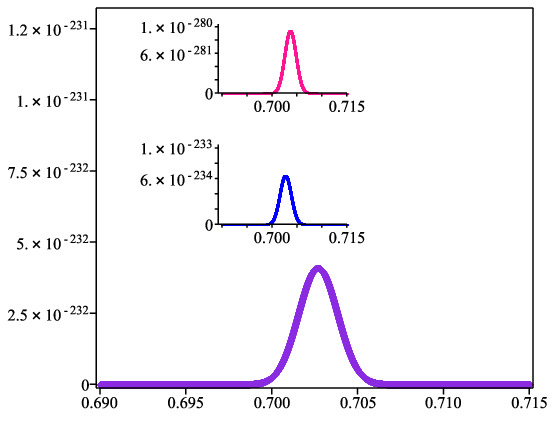}\\
\hspace*{0.045cm}\includegraphics[width=2cm, height=2.1cm]{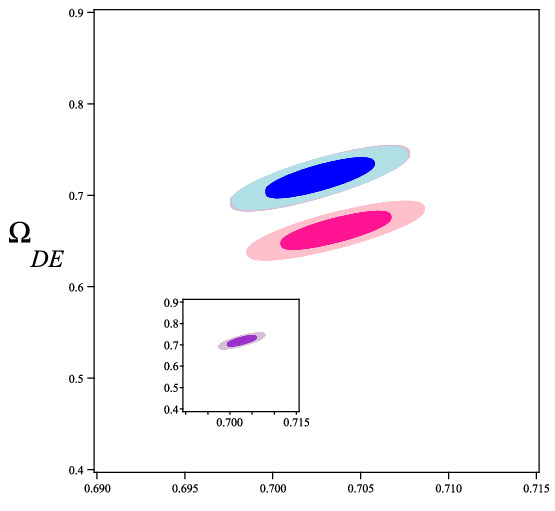} \includegraphics[width=2.1cm, height=2.09cm]{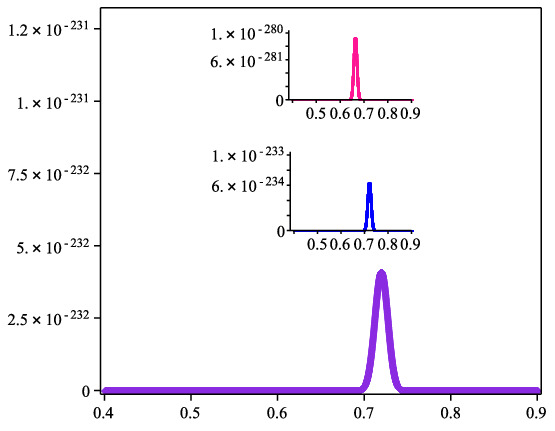}\\
\hspace*{0.06cm}\includegraphics[width=2.09cm,height=2.167cm]{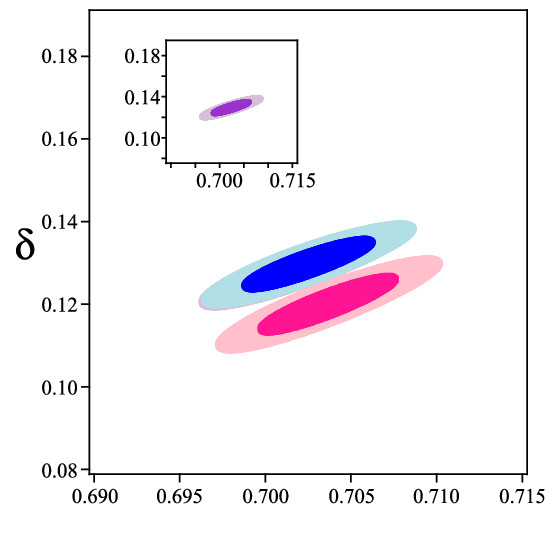}\hspace*{0.18cm}\includegraphics[width=1.91cm, height=2.16cm]{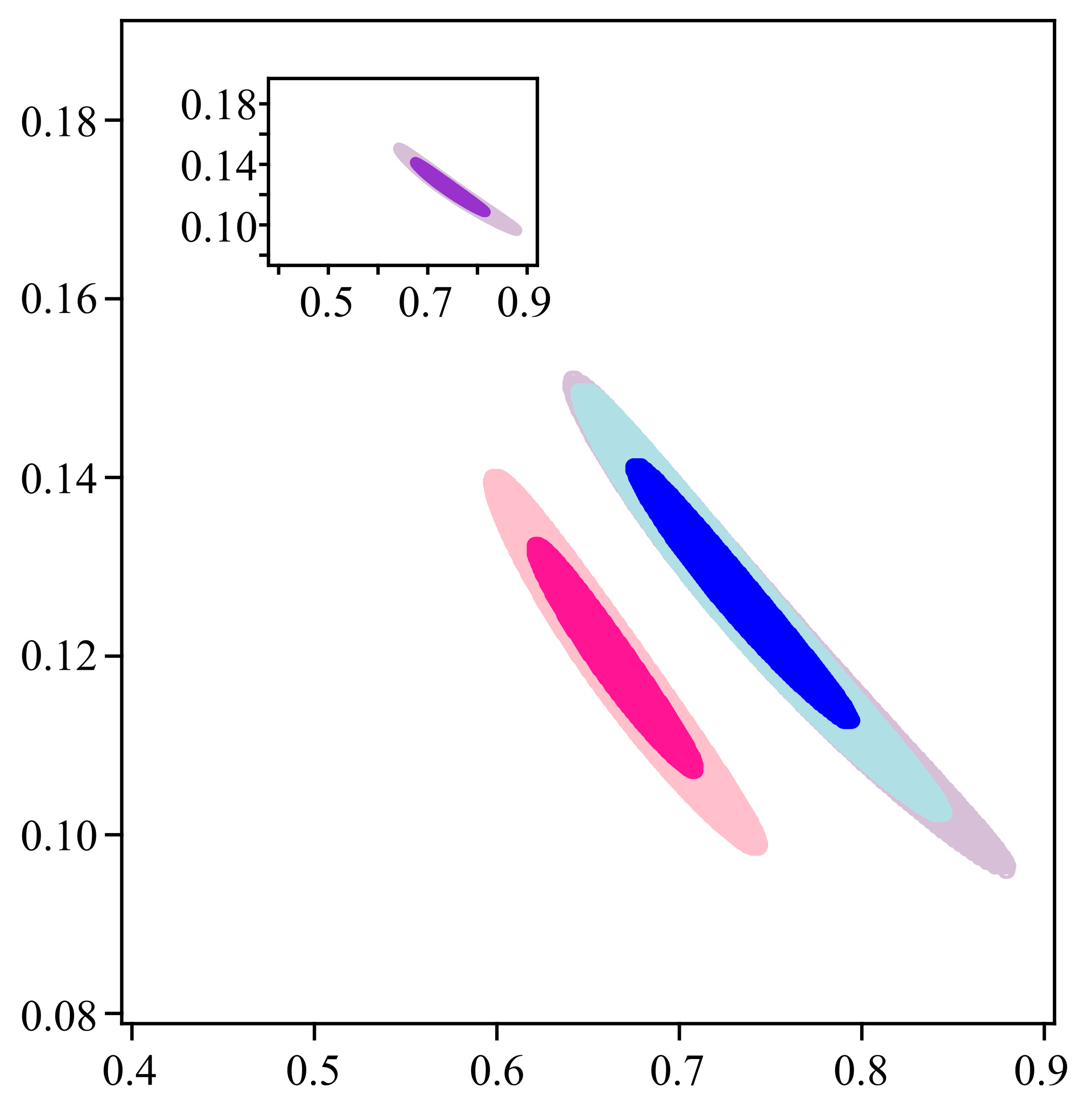}\includegraphics[width=2.15cm, height=2.16cm]{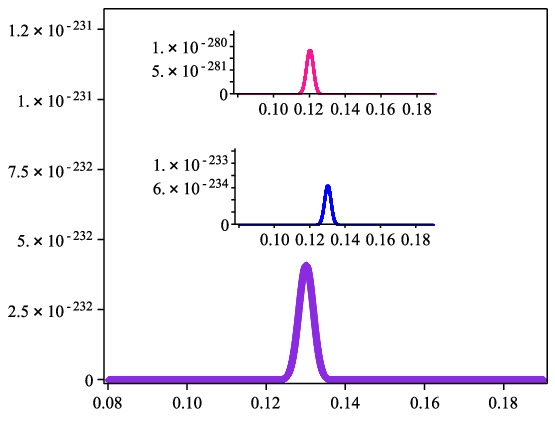}\\
\hspace*{0.02cm}\includegraphics[width=2.1cm, height=2.5cm]{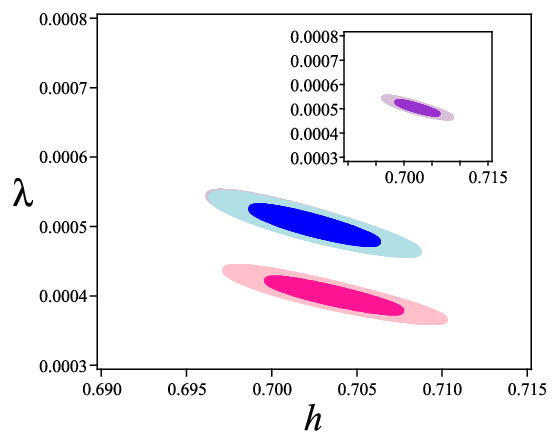}\hspace*{0.05cm}\includegraphics[width=2.03cm, height=2.49cm]{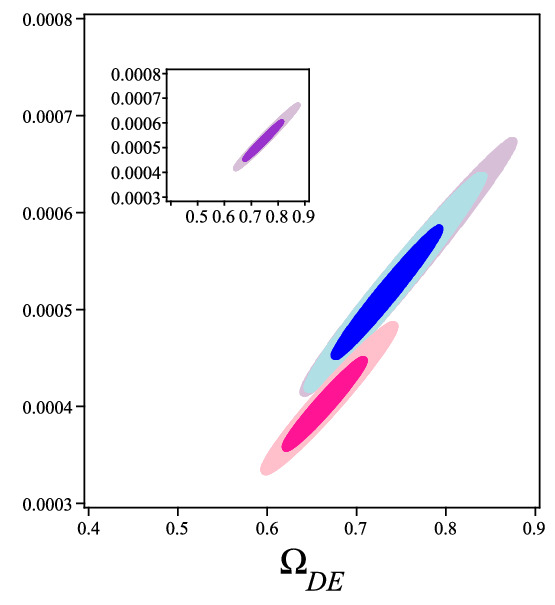}\hspace*{0.13cm}\includegraphics[width=2.04cm, height=2.48cm]{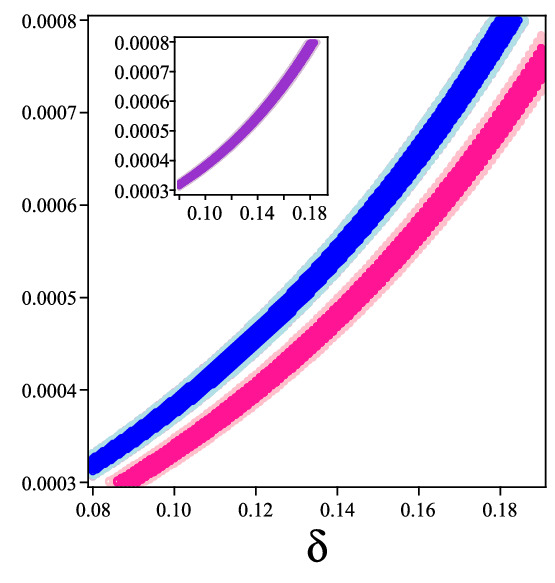}\includegraphics[width=2.34cm, height=2.415cm]{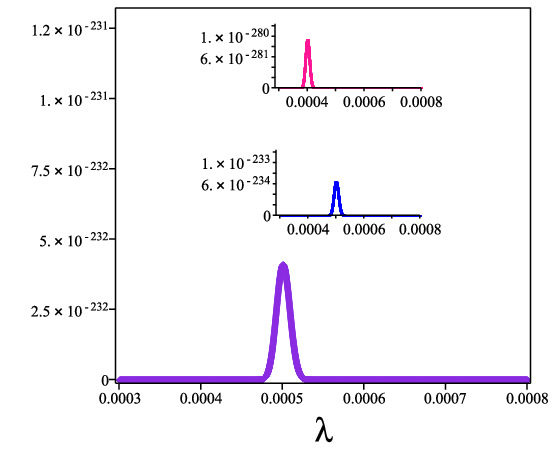}
\caption{2-dim contour levels and 1-dim likelihood distributions for the best-fit parameters of
 the noninteracting THDE with Ricci cutoff }
\label{fig:1}       
\end{figure*}

\begin{figure*}
\hspace*{0.12cm}\includegraphics[width=2.1cm, height=2.15cm]{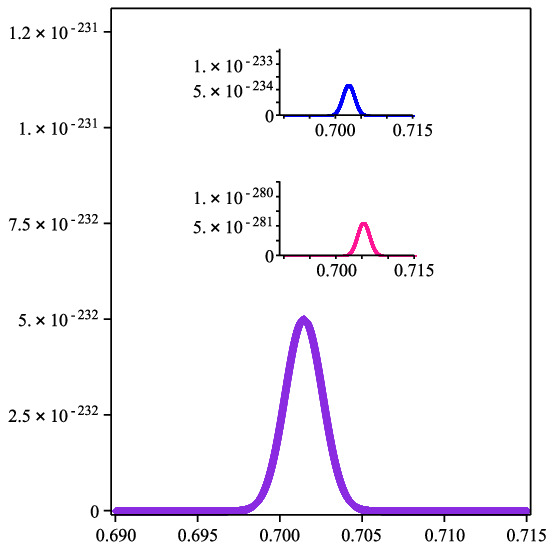}\\
\hspace*{0.12cm}\includegraphics[width=2.13cm, height=2.12cm]{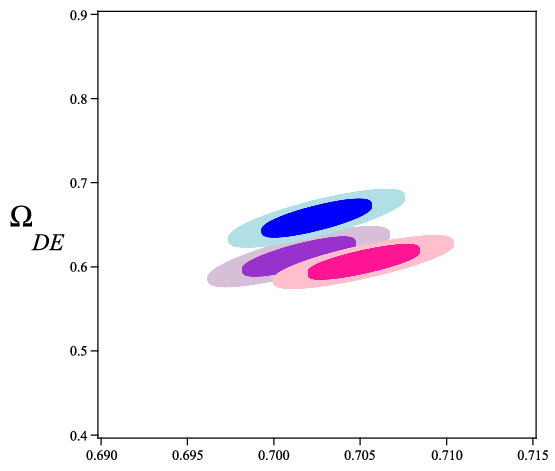}\hspace*{0.35cm}\includegraphics[width=2.28cm, height=2.12cm]{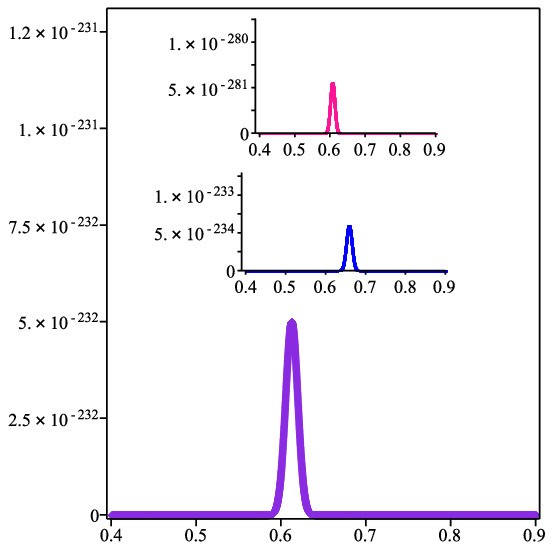}\\
\hspace*{0.312cm}\includegraphics[width=1.93cm, height=2.054cm]{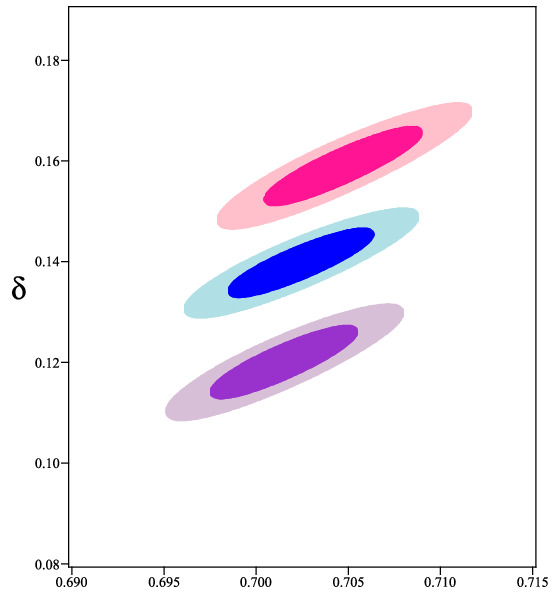}\hspace*{0.4cm}\includegraphics[width=2.0cm, height=2.106cm]{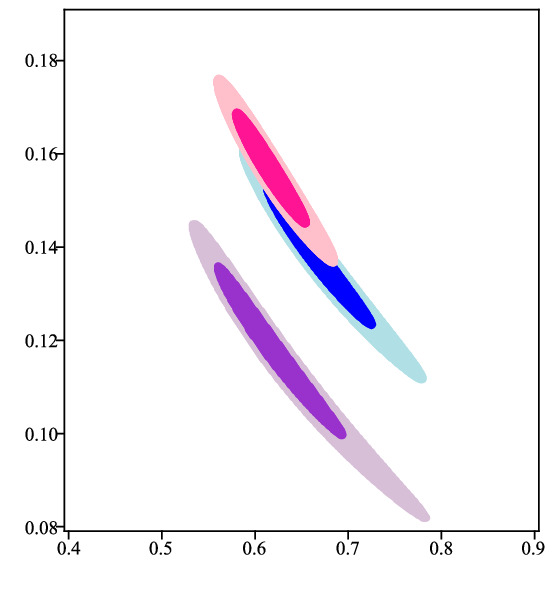}\includegraphics[width=2.255cm, height=2.118cm]{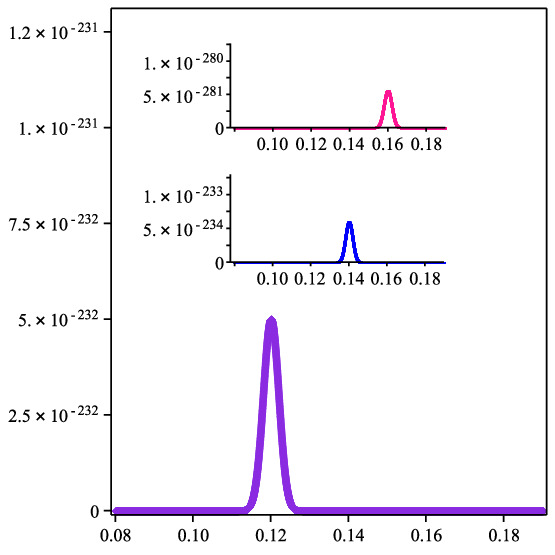}\\ \hspace*{0.145cm}\includegraphics[width=2.12cm, height=2.1cm]{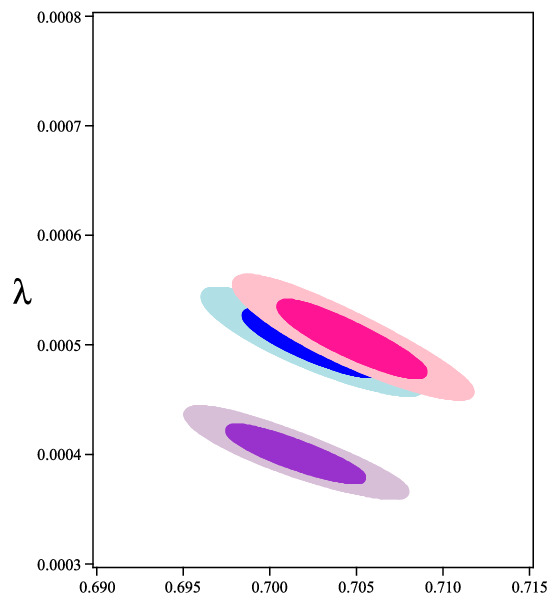}\hspace*{0.406cm}\includegraphics[width=1.96cm, height=2.16cm]{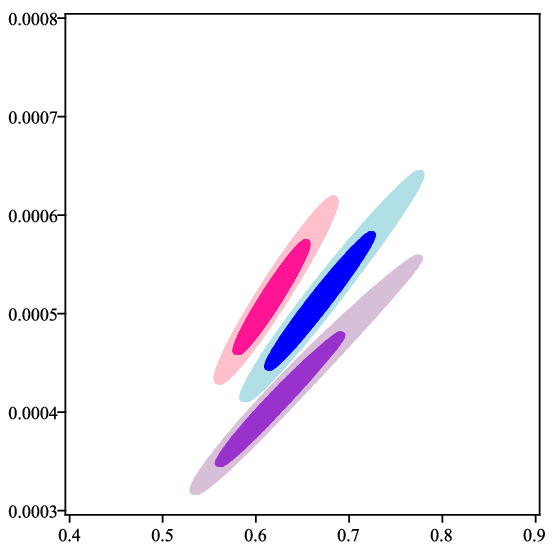}\hspace*{0.35cm}\includegraphics[width=2.0cm, height=2.16cm]{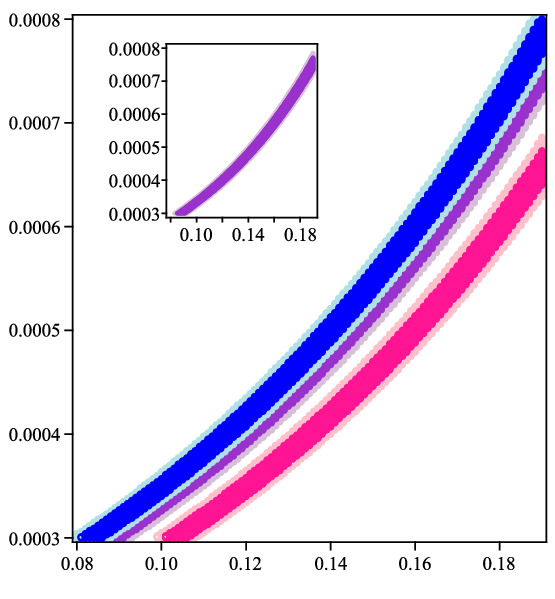}\includegraphics[width=2.5cm, height=2.15cm]{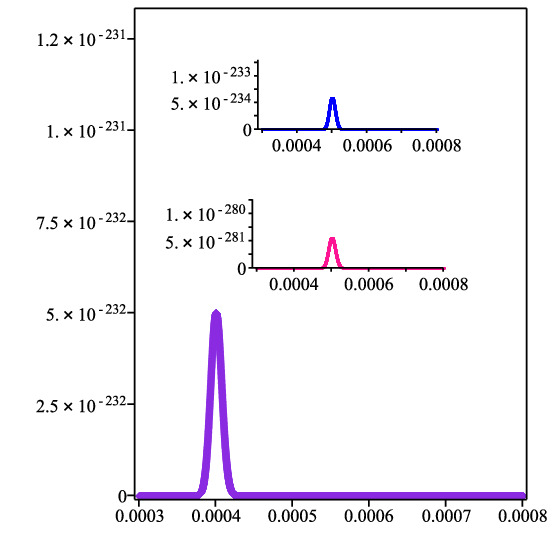}\\ \hspace*{0.12cm}\includegraphics[width=2.13cm, height=2.41cm]{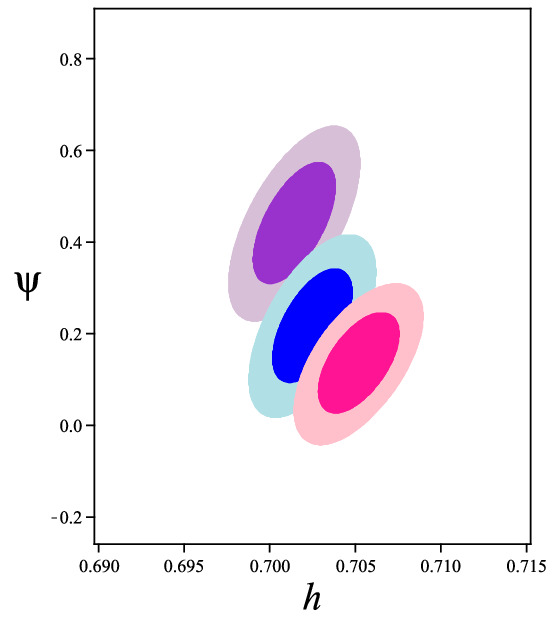}\hspace*{0.35cm}\includegraphics[width=2.02cm, height=2.40cm]{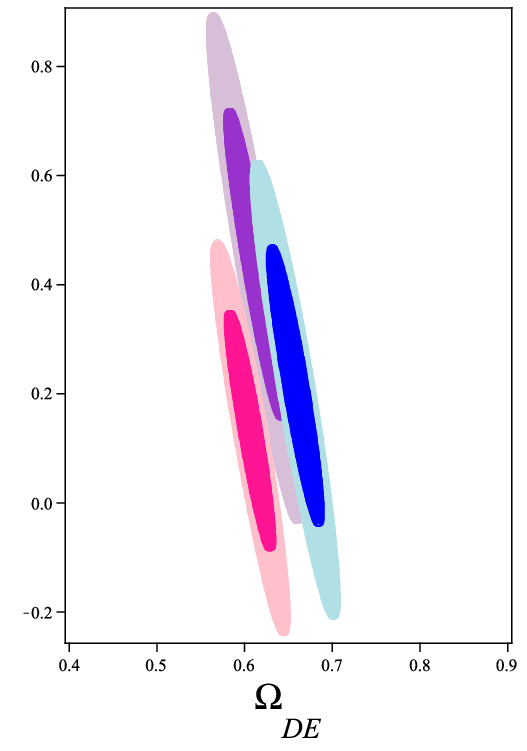}\hspace*{0.4cm}\includegraphics[width=1.9cm, height=2.42cm]{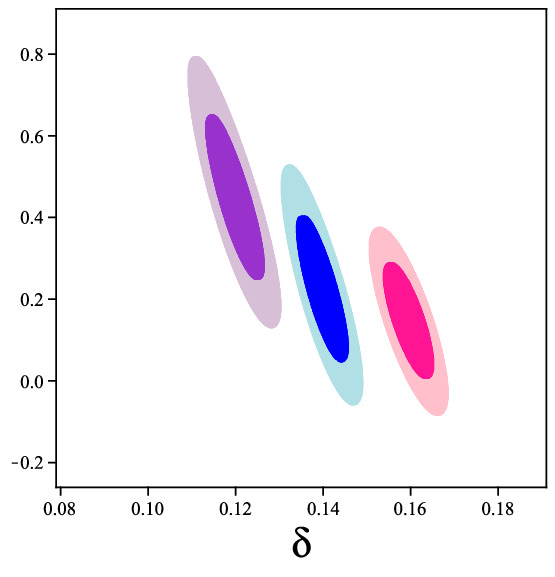}\hspace*{0.4cm}\includegraphics[width=2.1cm, height=2.44cm]{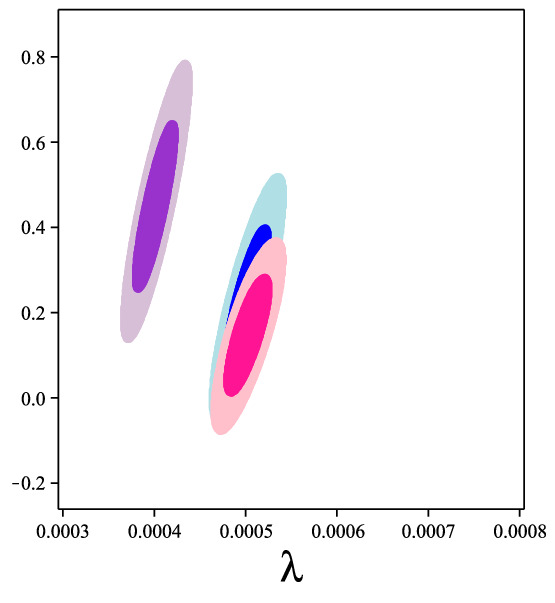}\includegraphics[width=2.43cm, height=2.45cm]{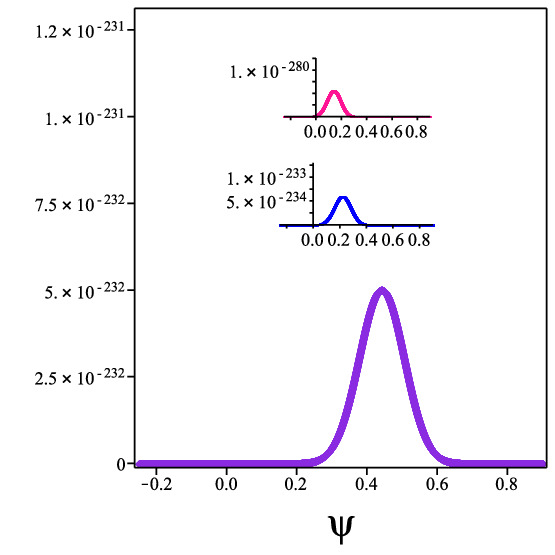}
\caption{2-dim contour levels and 1-dim likelihood distributions for the best-fit parameters of the nonlinear interacting THDE with Ricci cutoff }
\label{fig:1}       
\end{figure*}

\begin{table*}[t]
\caption{The best-fit parameters of the noninteracting THDE with Ricci cutoff }
\begin{tabular}{@{}lllllll|}
\hline
~Pars.&\hspace{0.0 cm}SNe Ia&\hspace{0.0 cm}SNe Ia+H(z)&\hspace{0.0 cm}SNe Ia+H(z)+GRB\\
\hline
\hline
~$h$~&\hspace{0.0 cm}$0.7027 ^{+0.0029+0.0047}_{-0.0028-0.0046}$&\hspace{0.0 cm}$0.7026^{+0.0029+0.0046}_{-0.0026-0.0043}$&\hspace{0.0 cm}$0.7036
^{+0.0029+0.0047}_{-0.0028-0.0045}$\\
\hline
~$\Omega_{DE}$~ &\hspace{0.0 cm}$0.719 ^{+0.019+0.031}_{-0.019-0.033}$&\hspace{0.0 cm}$0.719 ^{+0.019+0.030}_{-0.019-0.031}$&\hspace{0.0
cm}$0.662 ^{+0.017+0.027}_{-0.018-0.029}$\\
\hline
~$\delta$~&\hspace{0.0 cm}$0.13 ^{+0.005+0.009}_{-0.007-0.010}$&\hspace{0.0 cm}$0.13 ^{+0.005+0.009}_{-0.006-0.010}$&\hspace{0.0 cm}$0.12
^{+0.006+0.011}_{-0.007-0.011}$\\
\hline
~$\lambda$~&\hspace{0.0 cm}$0.0005 ^{+0.000030+0.000049}_{-0.000023-0.000039}$&\hspace{0.0 cm}$0.0005
^{+0.000028+0.000047}_{-0.000024-0.000040}$&\hspace{0.0 cm}$0.0004 ^{+0.000024+0.000041}_{-0.000023-0.000037}$\\
\hline
~$\chi^{2}_{min}$~&\hspace{0.0 cm}1065.581&\hspace{0.0 cm}1073.888&\hspace{0.0 cm}1289.572\\
\hline
\end{tabular}
\end{table*}

\begin{table*}[t]
\caption{The best-fit parameters of the nonlinear interacting THDE with Ricci cutoff }
\begin{tabular}{@{}lllllll|}
\hline
~Pars.&\hspace{0.0 cm}SNe Ia&\hspace{0.0 cm}SNe Ia+H(z)&\hspace{0.0 cm}SNe Ia+H(z)+GRB\\
\hline
\hline
~$h$~&\hspace{0.0 cm}$0.7014 ^{+0.0031+0.0049}_{-0.0028-0.0047}$&\hspace{0.0 cm}$0.7025^{+0.0029+0.0046}_{-0.0027-0.0046}$&\hspace{0.0 cm}$0.7052
^{+0.0030+0.0047}_{-0.0029-0.0048}$\\
\hline
~$\Omega_{DE}$~ &\hspace{0.0 cm}$0.612 ^{+0.019+0.030}_{-0.019-0.032}$&\hspace{0.0 cm}$0.658 ^{+0.018+0.029}_{-0.019-0.030}$&\hspace{0.0
cm}$0.606 ^{+0.016+0.026}_{-0.017-0.027}$\\
\hline
~$\delta$~&\hspace{0.0 cm}$0.12 ^{+0.007+0.011}_{-0.007-0.011}$&\hspace{0.0 cm}$0.14 ^{+0.006+0.010}_{-0.006-0.010}$&\hspace{0.0 cm}$0.16
^{+0.006+0.010}_{-0.008-0.013}$\\
\hline
~$\lambda$~&\hspace{0.0 cm}$0.0004 ^{+0.000024+0.000040}_{-0.000024-0.000037}$&\hspace{0.0 cm}$0.0005
^{+0.000029+0.000046}_{-0.000025-0.000041}$&\hspace{0.0 cm}$0.0005 ^{+0.000037+0.000059}_{-0.000027-0.000046}$\\
\hline
~$\psi$~ &\hspace{0.0 cm}$0.441 ^{+0.121+0.191}_{-0.118-0.199}$&\hspace{0.0 cm}$0.218
^{+0.115+0.190}_{-0.119-0.195}$&\hspace{0.0 cm}$0.138 ^{+0.088+0.140}_{-0.095-0.160}$\\
\hline
~$\chi^{2}_{min}$~&\hspace{0.0 cm}1065.180&\hspace{0.0 cm}1074.033&\hspace{0.0 cm}1290.611\\
\hline
\end{tabular}
\end{table*}

The results of the best fit values of the Tsallis holographic dark energy (THDE) and nonlinear interacting Tsallis holographic dark energy (NITHDE) models have been prepared in Tables 2 and 3. We have plotted the confidence
levels (68.3$\%$ and 95.4$\%$) and 1-dim likelihood graphs in Figs. 1 and 2.
\\\\\\\\\\\\\\\\\\\\\\\\\\\\\\\\\\\\\\\\
\\\\\\\\\\\\\\\\\\\\\\\\\\\\\\\\\\\\\\\\\\\\

\section{$H(z)$ plane and turning point feature}\label{sec4}

The $H(z)$ plane corresponds to the comparison of the model (or theory) versus observational data points during the
cosmic time.
Apart from this, recently, some authors have examined an exotic feature on the Hubble plot (with respect to redshift)
called a turning point. \cite*{bib55} have searched for the turning point of the HDE model considering the arbitrary
and best-fit values with the future event horizon. The results for the $\Omega_{DE0}=0.7$, $H_{0}=100$, and $c<1$
have identified that the turning point for $c<1$ is inevitable.
However, the measurements of the observational data, including CMB+BAO and CMB+BAO+SNe Ia, indicate that CMB+BAO data
has a turning point at $z\simeq 0.04$, while CMB+BAO+SNe Ia leads to a turning point at $z\simeq -0.1$. Investigating
the turning point in the Barrow holographic dark energy (BHDE) with Hubble and FE cutoffs has been worked by
\cite*{bib56}. The authors have discussed the BHDEF model with different $C$ and $\Delta$. The results illustrate that
increasing the $C$ (or decreasing the $\Delta$) parameter causes the turning point to move to the low $z$, while it
vanishes at $C\geq 4.45$ (or $\Delta\leq 0.04$). Furthermore, authors have mentioned that for $C=
3.421^{+1.753}_{-1.611}$ and $\Delta= 0.094^{+0.094}_{-0.101}$ measured by \cite*{bib57} for H(z)+SNe Ia data, the
turning point does not exist. Thus, they have concluded that the BHDE models will not have a turning point. Finally, \cite*{bib43} has Observed a turning point in the Tsallis agegraphic dark energy (with $\rho _{DE}=m
T^{2\delta -4}$) in DGP braneworld cosmology with a linear interaction term $Q=3b^{2}H(\rho _{DE}+\rho _{m})$. She has
measured $\Omega _{DE}(z=0)=0.599$, $\delta=12.7$, $m=3.0$, $b=0.28$, and $\Omega _{r_{c}}=0.0018$ dimensionless
parameters using the 30 Hubble dataset and has shown the model has a turning point in the range of $-1<z<-0.5$.

\begin{figure}[h]%
\centering
\includegraphics[width=0.6\textwidth]{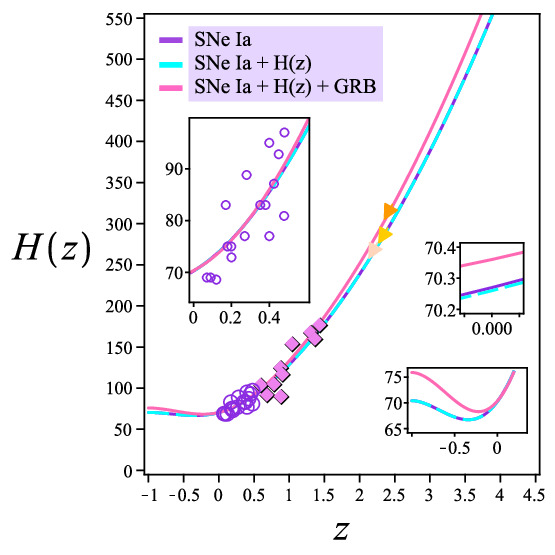}\\
\includegraphics[width=0.6\textwidth]{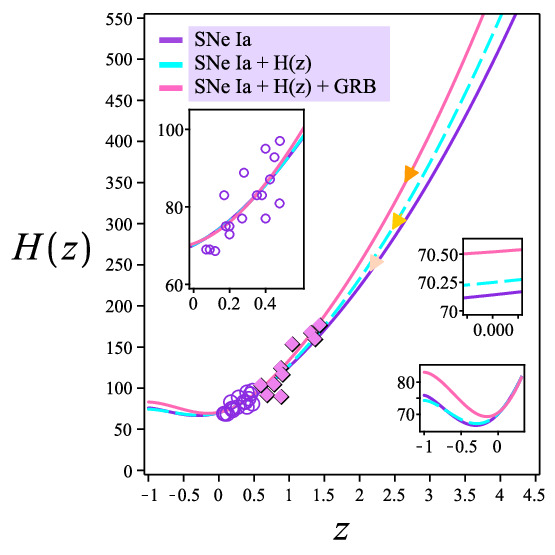}
 \caption{The Hubble parameter versus redshift in comparison with the observational data  }
 \label{fig:3}       
\end{figure}

Now, to find the turning point in our models, we can observe Fig. 3. As it is obvious, there is a turning point at $(z,H_{0})=(-0.343,66.743)$, $(z,H_{0})=(-0.343,66.739)$, and $(z,H_{0})=(-0.229,68.311)$ for the THDE with Ricci cutoff with SNe Ia, SNe Ia+H(z), and SNe Ia+H(z)+GRB data, respectively. This point will occur at $(z,H_{0})=(-0.302,66.584)$, $(z,H_{0})=(-0.288,67.154)$, and $(z,H_{0})=(-0.148,69.317)$ for the (nonlinear) interacting THDE model with the SNe Ia, SNe Ia+H(z), and SNe Ia+H(z)+GRB datasets.
\\\\\\\\\\\\\\\\\\\\\\\\\\\\\\\\\\\\\\\\\\\\\\\\\\\\\\\\

\section{Some features of the model}\label{sec5}

This section is allocated to analyze the evolution of dark energy EoS, effective EoS, and
deceleration parameters against the redshift for the fitted parameters of the models.

The dark energy and the effective equation of state parameters for the THDE with Ricci cutoff with the consideration
of $Q=0$ and $Q\neq0$ can be given by
\begin{eqnarray}
&&w_{DE}=-1-\frac{1-\Omega_{DE}}{\Omega_{DE}}\nonumber\\&&-\frac{2}{3\Omega_{DE}}(\frac{(3H^{2}\Omega_{DE})^\frac{1}{2-\delta}}{H^{2}\lambda^{\frac{1}{2-\delta}}}-2).
\end{eqnarray}

and

\begin{eqnarray}
&&w_{eff}=\frac{P_{eff}}{\rho_{eff}}=\frac{P_{m}+P_{DE}}{\rho_{m}+\rho_{DE}}=\Omega_{DE} w_{DE}.
\end{eqnarray}

The deceleration parameter will be

\begin{eqnarray}
&&q=-1-\frac{\frac{dH}{dX}}{H}=\nonumber\\&&\frac{1}{H^{2} \lambda^\frac{1}{2-\delta}}[H^{2} \lambda^\frac{1}{2-\delta}-(3 H^{2}\Omega_{DE})^\frac{1}{2-\delta}].
\end{eqnarray}

The results of $w_{DE}$, $w_{eff}$, and $q$, for both THDE models with Ricci cutoff with $Q=0$ and $Q\neq0$, are plotted in Figures 4-6.

From the $w_{DE}$ plot against redshift, we can analyze the present value of $w_{DE}$ ($w_{DE0}$) and survey the quintessential, phantom, or cosmological behavior of the DE during cosmic evolution. So, As we see in Fig. 4 (up panel), in the THDE with the Ricci horizon without interaction term, the $w_{DE0}=-1.094$, $w_{DE0}=-1.095$, and $w_{DE0}=-1.232$ for SNe Ia, SNe Ia+H(z), and SNe Ia+H(z)+GRB samples, respectively. But, for the nonlinear interacting THDE with the Ricci cutoff, the $w_{DE0}$ is -1.262 (for SNe Ia), -1.198 (for SNe Ia+H(z)), and -1.395 (for SNe Ia+H(z)+GRB). Also, from Fig. 4, we can observe that for the THDE and nonlinear interacting THDE with Ricci cutoff as IR, the DE behaves as quintessence in the past and phantom at the current epoch. Then, dark energy will approach the $\Lambda$ state at a late time.

Our results of the $w_{DE0}$ considering the THDE and nonlinear interacting THDE with Ricci horizon as IR cutoff lies between $-1.03\leq w_{DE0}\leq-1.55$ obtained by \cite*{bib58} for the $TT+Low E$, $TT^{clean}+Low E$, $TT,TE,EE+Low E$, $TT,TE,EE+Low E+Lensing$, and $TT,TE,EE+Low E+Lensing+BAO$ datasets.

Following the figures, we can recognize the phase of the Universe and the current value of the $w_{eff}$ in Fig. 5.
From this plot, both the THDE with noninteraction and (nonlinear) interaction terms with Ricci cutoff can experience the quintessence era at high and current redshifts. The models will enter the phantom region and then turn to move -1 in the future era. Continuing to analyze Figure 5, the $w_{eff0}$ is -0.787 (-0.772) for SNe Ia, -0.787 (-0.788) for SNe Ia+H(z), and -0.816 (-0.845) for SNe Ia+H(z)+GRB in the noninteracting THDE (nonlinear interacting THDE) model.

\begin{figure}[h]%
\center
\includegraphics[width=0.6\textwidth]{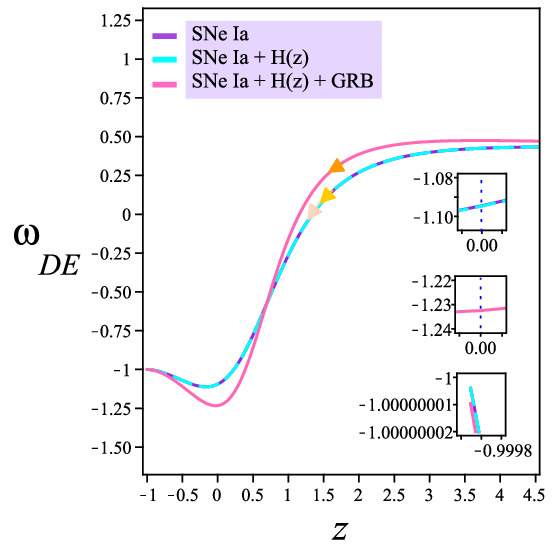}\\
\includegraphics[width=0.6\textwidth]{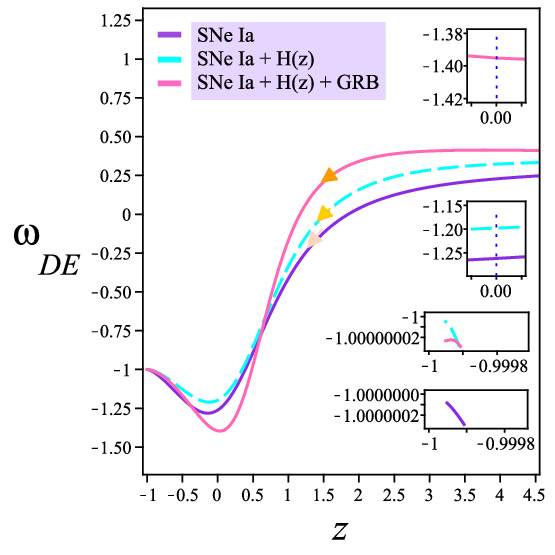}
 \caption{The evolution of $w_{DE}$ parameter versus redshift for the noninteracting THDE with Ricci cutoff (the up plot)
  and nonlinear interacting THDE with Ricci cutoff (the bottom plot)}
 \label{fig:4}       
\end{figure}

\begin{figure}[h]%
\center
\includegraphics[width=0.6\textwidth]{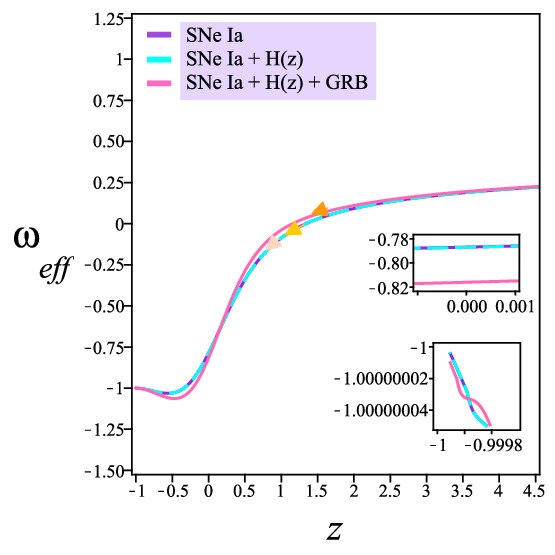}\\
\includegraphics[width=0.6\textwidth]{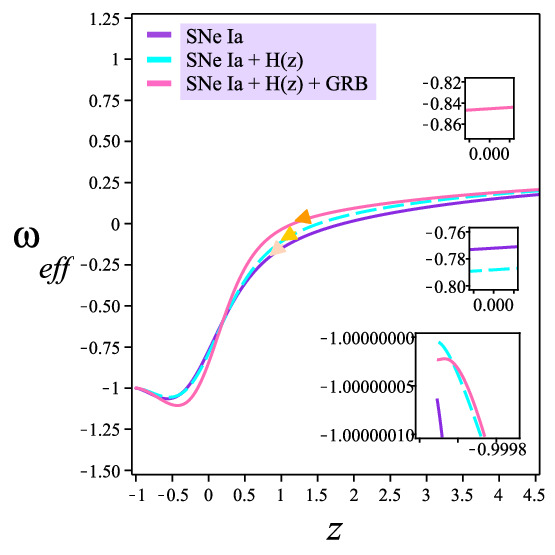}
 \caption{The evolution of $w_{eff}$ parameter versus redshift for the noninteracting THDE with Ricci cutoff (the up plot)
  and nonlinear interacting THDE with Ricci cutoff (the bottom plot)
 }
 \label{fig:5}       
\end{figure}

In addition to plotting the $w_{DE}$ and $w_{eff}$ parameters versus redshift, we have depicted the $q$ parameter against redshift in Fig. 6. From this figure, today's value of the deceleration parameter can be recognized. Moreover, when the curve of $q$ reaches $q=0$, we can find the transition redshift from matter-dominated to dark energy-dominated phase (or starting point of the acceleration phase of the Universe). So, we have found the value of $z_{tr}$ in our models directly from the plots in Fig. 6 measuring the redshift at $q=0$.

So, for the THDE with Ricci cutoff with $Q=0$ in the up plot of Fig. 6, we have $q_{0}$= -0.680, -0.681, and -0.724 and $z_{tr}$= 0.510, 0.510, and 0.446 for the SNe Ia, SNe Ia+H(z), and SNe Ia+H(z)+GRB samples, respectively.
Furthermore, the bottom plot of Fig. 6 is depicted to show the evolution of $q$ versus redshift for the nonlinear interacting THDE with Ricci as IR. From this panel, the $q_{0}$ and $z_{tr}$ will be obtained as -0.658
and 0.552 for SNe Ia, -0.682 and 0.517 for SNe Ia+H(z), and -0.768 and 0.427 for SNe Ia+H(z)+GRB sample.

Recently, \cite*{bib59} have constrained the THDE with GO cutoff with linear interaction term $Q=3bH\rho_{m}$ using SNIa, OHD, SNIa+OHD, SNIa+OHD+CMB, SNIa+OHD+BAO, and SNIa+OHD+CMB+BAO data. The authors have studied the cosmological parameters, dynamical analysis, and thermodynamics of the model. However, here we only mention the results of $q_{0}$ and $z_{tr}$ calculated from the $q$ plot against redshift for two datasets. The authors have found the $q_{0}=-0.602$ and $z_{tr}=0.799$ for SNIa+OHD+CMB data and $q_{0}=-0.594$ and $z_{tr}=0.763$ for SNIa+OHD+CMB+BAO data.

In another work, \cite*{bib25} has constrained the THDE with the Hubble and future event cutoffs with $Q=0$ and $Q=3bH\rho_{m}$ for BAO+CMB, BAO+CMB+SNe Ia, BAO+CMB+SNe Ia+GRB, and BAO+CMB+SNe Ia+GRB+H samples. Here, the author has fitted $z_{tr}$ and has shown $z_{tr}$ for the THDE with Hubble (future event) horizon with $0.604\leq z_{tr}\leq0.677$ ($0.502\leq z_{tr}\leq0.572$), while for the linear interacting THDE with Hubble (future event) cutoff we have $0.503\leq z_{tr}\leq0.634$ ($0.541\leq z_{tr}\leq0.649$).\\\\\\\\\\\\\\\\\\\\\\\\\\\\\\\\\\\\\\\\\\\\\\\\\\

\begin{figure}[h]%
\center
\includegraphics[width=0.6\textwidth]{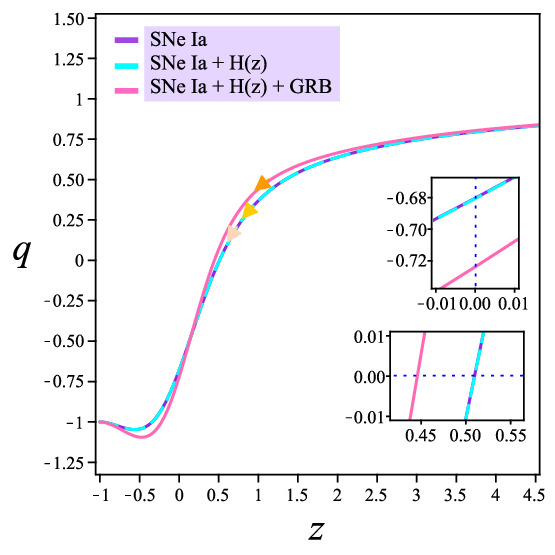}\\
\includegraphics[width=0.6\textwidth]{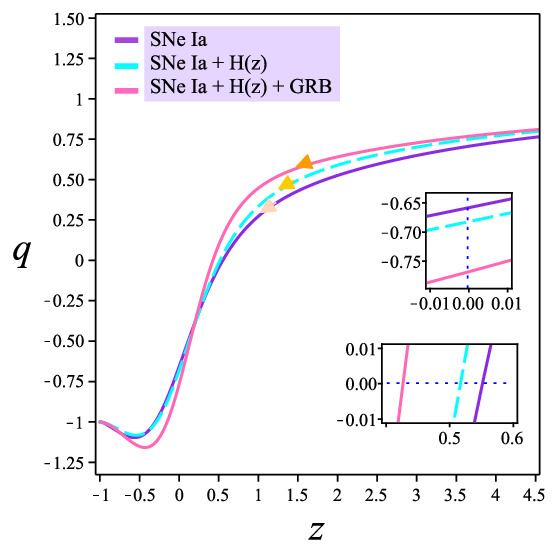}
 \caption{The evolution of $q$ parameter versus redshift for the noninteracting THDE with Ricci cutoff (the up plot)
  and nonlinear interacting THDE with Ricci cutoff (the bottom plot)}
 \label{fig:6}       
\end{figure}

\section{Stability}\label{sec6}

The sign of the squared sound speed parameter can be employed to find the stability of the models. The positive value of $v_{s}^{2}$ shows sustainability, whereas the negative sign describes the unsustainability of the models. The $v_{s}^{2}$ is represented by

\begin{eqnarray}\label{defin1}
v_{s}^{2}=\frac{dP_{DE}}{d\rho_{DE}}=H\frac{\rho_{DE}}{\dot{\rho}_{DE}} \omega_{DE}^{'}+\omega_{DE}.\nonumber
\end{eqnarray}

For the THDE with Ricci cutoff with the noninteraction term,

\begin{eqnarray}
&&v_{s}^{2}=(\frac{\Omega_{DE}}{2\frac{1}{H}\frac{dH}{dX}+3(1-\Omega_{DE})})\omega_{DE}^{'}+\omega_{DE},\nonumber
\end{eqnarray}

 and,

\begin{eqnarray}
&&v_{s}^{2}=(\frac{\Omega_{DE}}{2\frac{1}{H}\frac{dH}{dX}-3\psi\Omega_{DE}(1-\Omega_{DE})+3(1-\Omega_{DE})})\omega_{DE}^{'}
\nonumber\\&&+\omega_{DE},\nonumber
\end{eqnarray}

for nonlinear interaction case.

Which $\omega^{'}_{DE}$ is

\begin{eqnarray}
&&\omega^{'}_{DE}=-\frac{4(3H^{2}\Omega_{DE})^\frac{1}{2-\delta}}{3(2-\delta)H\Omega_{DE}\lambda^\frac{1}{2-\delta}}\frac{\frac{dH}{dX}}{H}
\nonumber\\
&&-\frac{2(3H^{2}\Omega_{DE})^\frac{1}{2-\delta}}{3\lambda^\frac{1}{2-\delta}(2-\delta)H\Omega_{DE}^{2}}\frac{d\Omega_{DE}}{dX}
\nonumber\\
&&+\frac{2(3H^{2}\Omega_{DE})^\frac{1}{2-\delta}}{3\lambda^\frac{1}{2-\delta}H\Omega_{DE}^{2}}\frac{d\Omega_{DE}}{dX}
\nonumber\\
&&+\frac{4(3H^{2}\Omega_{DE})^\frac{1}{2-\delta}}{3\lambda^\frac{1}{2-\delta}H\Omega_{DE}}\frac{\frac{dH}{dX}}{H}-\frac{\frac{Hd\Omega_{DE}}{dX}}{3\Omega_{DE}^{2}}.
\end{eqnarray}

\begin{figure}[h]%
\centering
\includegraphics[width=0.6\textwidth]{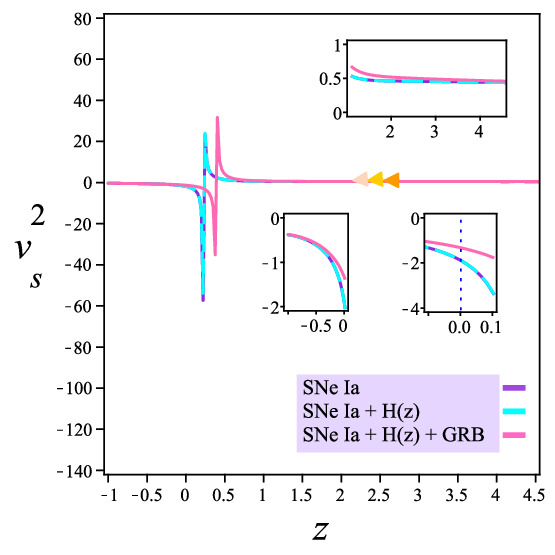}\\
\includegraphics[width=0.6\textwidth]{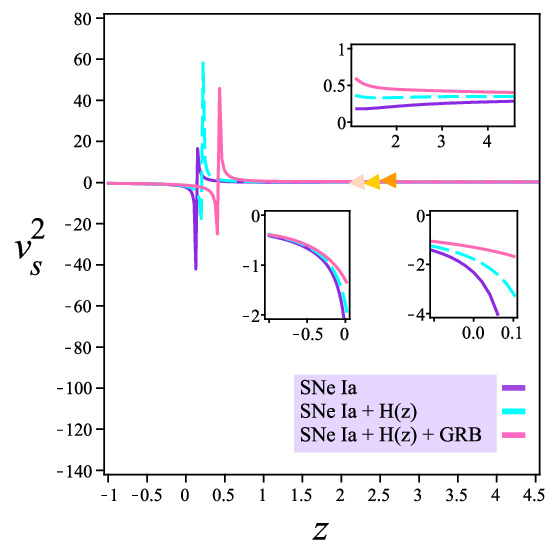}
 \caption{The evolution of $v_{s}^{2}$ versus redshift for the noninteracting THDE with Ricci cutoff (the up plot)
  and nonlinear interacting THDE with Ricci cutoff (the bottom plot) }
 \label{fig:7}       
\end{figure}

Now, we would like to use the evolution of the square sound of speed to identify the stability ($v_{s}^{2}>0$) or
instability ($v_{s}^{2}<0$) of our models during the evolution of the Universe (see Fig. 7).
As it is manifest, both the THDE models with a noninteraction and nonlinear interaction terms for the SNe Ia, SNe Ia+H(z), and SNe Ia+H(z)+GRB data display the stable models at the early time and unstable models in the current and future time.

Now, what we like to do here is to compare the THDE with the Ricci cutoff between this manuscript and the \cite*{bib24} project in terms of the $v_{s}^{2}$ (see Table 4). From this Table, we observe that the noninteracting THDE with Ricci horizon in both works shows stability in the past and instability in the current and future eras.
 Although the linear interacting THDE with Ricci as L does not provide a stable model, the THDE with Ricci horizon regarding the nonlinear interaction term displays a sustainable Universe at the early epoch. As more comparison, in Table 5, we are comparing the results of the $v_{s}^{2}$ between the THDE with different IR cutoffs (Ricci, Hubble, future event, GO, and particle cutoffs).

\begin{table*}[t]
\normalsize
\caption{The comparison of $v_{s}^{2}$ for the THDE model with Ricci as IR cutoff.
Here, ITHDE and NITHDE are the linear and nonlinear interacting THDE with Ricci horizon, respectively}
\begin{tabular}{|@{}l|l|l|l|l|l|}
\hline
~Model& Past&Present&Future&Ref.\\
\hline
\hline
~THDE (Ricci cutoff)& $>0$ &$<0$ &$<0$& This project\\
\hline
~NITHDE (Ricci cutoff)& $>0$ &$<0$ &$<0$&This project \\
\hline
~THDE (Ricci cutoff)& $>0$ &$<0$ &$<0$& \cite*{bib24}\\
\hline
~ITHDE (Ricci cutoff)& $<0$ &$<0$ &$<0$& \cite*{bib24} \\
\hline
\end{tabular}
\end{table*}

\begin{table*}[t]
\normalsize
\caption{The comparison of $v_{s}^{2}$ for the THDE model with different IR cutoffs.
Here, ITHDE and NITHDE are the linear and nonlinear interacting THDE models, respectively}
\begin{tabular}{|@{}l|l|l|l|l|l|}
\hline
~Model& Past&Present&Future&Ref.\\
\hline
\hline
~THDE (Ricci cutoff)& $>0$ &$<0$ &$<0$& This project\\
\hline
~NITHDE (Ricci cutoff)& $>0$ &$<0$ &$<0$&This project \\
\hline
~THDE (Ricci cutoff)& $>0$ &$<0$ &$<0$& \cite*{bib24}\\
\hline
~ITHDE (Ricci cutoff)& $<0$ &$<0$ &$<0$& \cite*{bib24} \\
\hline
~THDE (GO cutoff)& $>0$ &$<0$ &$<0$& \cite*{bib24}\\
\hline
~ITHDE (GO cutoff)& $>0$ &$<0$ &$<0$& \cite*{bib24} \\
\hline
~THDE (particle cutoff)& $<0$ &$<0$ &$<0$& \cite*{bib24}\\
\hline
~ITHDE (particle cutoff)& $>0$ &$<0$ &$<0$& \cite*{bib24} \\
\hline
~THDE (Hubble cutoff)&  $<0$ & $<0$& $<0$&\cite*{bib25}\\
\hline
~ITHDE (Hubble cutoff)&  $<0$ & $<0$& $<0$&\cite*{bib25}\\
\hline
~ITHDE (Hubble cutoff)&  $<0$ & $<0$& $<0$&\cite*{bib24}\\
\hline
~THDE (future event cutoff)&$<0$ & $<0$&$>0$&\cite*{bib25} \\
\hline
~ITHDE (future event cutoff)&$<0$ & $<0$&$>0$&\cite*{bib25} \\
\hline
\end{tabular}
\end{table*}

From Table 5, it can be recognized that the THDE and linear THDE model with Hubble horizon and the THDE with particle cutoff will not behave stably during the cosmic time. But, we see the sustainability of the THDE and linear THDE with future event horizons at the late time. Amongst the different mentioned models in Table 5, the THDE and nonlinear THDE with Ricci horizon, THDE and linear THDE with GO cutoff, and linear THDE with particle horizon can show stability at the past epoch.

\section{Jerk and $OM$ diagnostic tool}\label{sec7}

What we tend to investigate in this section is to survey the evolution of the Jerk parameter and the $OM$ diagnostic to discriminate between the $\Lambda CDM$ and the noninteracting and nonlinear interacting THDE with Ricci horizon as IR cutoff.\\\\

\subsection{The Jerk parameter}\label{subsec1}

\begin{figure}[h]%
\centering
\includegraphics[width=0.6\textwidth]{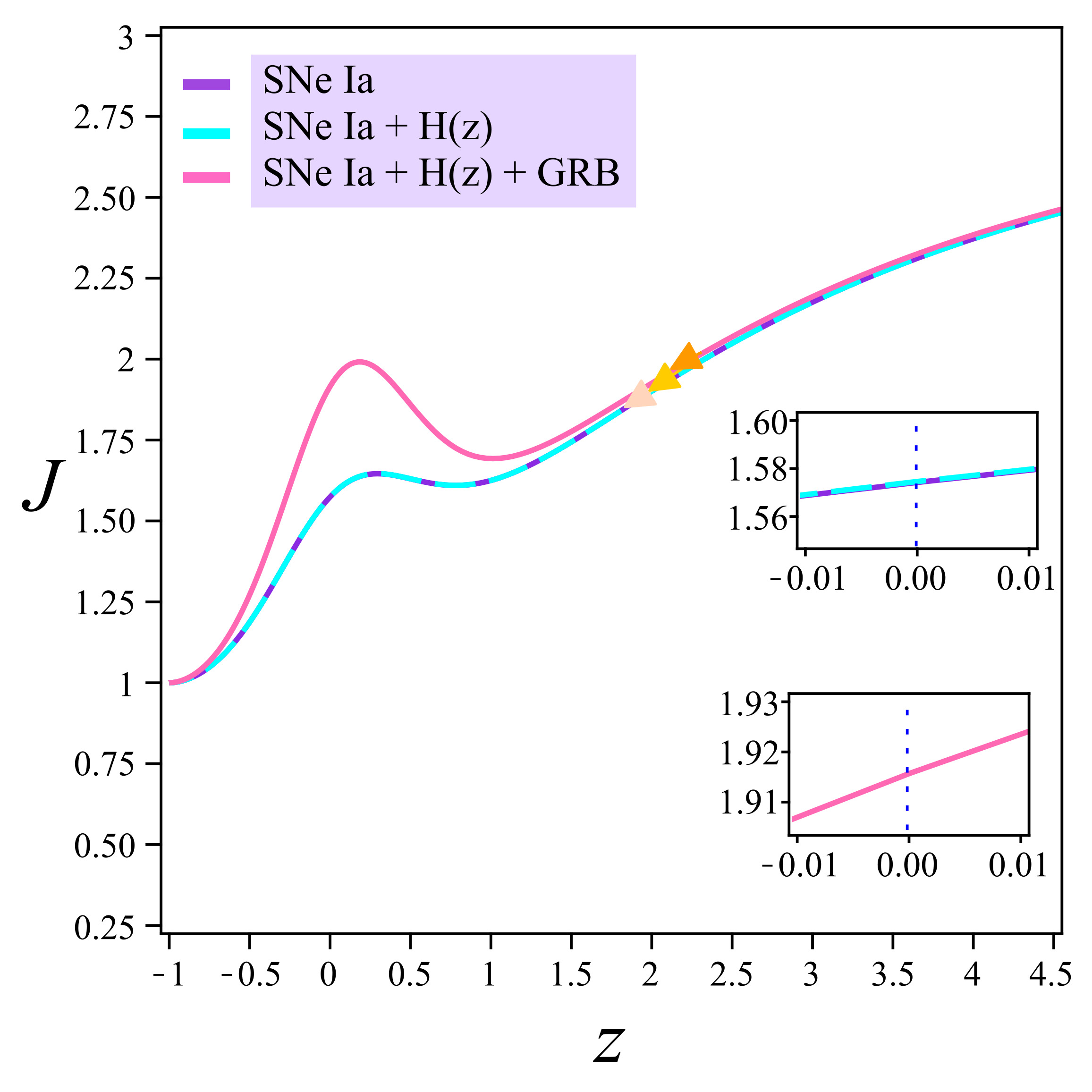}\\
\includegraphics[width=0.6\textwidth]{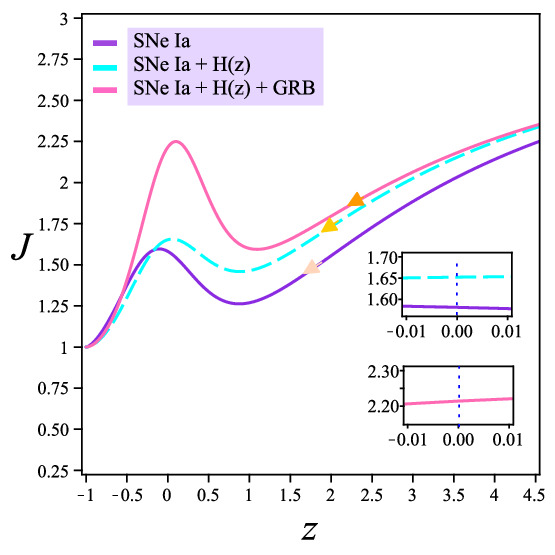}
 \caption{The evolution of $Jerk$ parameter versus redshift for the noninteracting THDE with Ricci cutoff (the up plot)
  and nonlinear interacting THDE with Ricci cutoff (the bottom plot)}
 \label{fig:8}       
\end{figure}

The $Jerk$ parameter ($J$) is mainly a way to compare the models with the $\Lambda CDM$ model characterized by $J=1$ in the flat Universe. Also, the $Jerk$ parameter
with a positive value can describe the acceleration phase of the Universe. The $J$ is expressed by

 \begin{eqnarray}\nonumber
J= \frac{\ddot{H}}{H^{3}}-3q-2.
\end{eqnarray}

So,

\begin{eqnarray}
&&J=\frac{\frac{dH}{dX}}{H}(-4+\frac{2(3H^{2}\Omega_{DE})^\frac{1}{2-\delta}}{H^{2}(2-\delta)\lambda^{\frac{1}{2-\delta}}})
\nonumber\\
&&+\frac{(3H^{2}\Omega_{DE})^\frac{1}{2-\delta}}{H^{2}(2-\delta)\Omega_{DE}\lambda^{\frac{1}{2-\delta}}}\frac{d\Omega_{DE}}{dX}
\nonumber\\
&&-3q-2.
\end{eqnarray}

The Jerk parameter of the THDE without interaction and with nonlinear interaction term with Ricci as IR L (see relation 16) have been graphed in Fig
8. As it is explicit, our models deviate from the $\Lambda CDM$ model from the past to the present. However, the models will approach the $\Lambda CDM$ model in the future. In addition, from figure 8, the present value of $J$  ($J_{0}$) is 1.574 (SNe Ia), 1.575 (SNe Ia+H(z)), and 1.916 (SNe Ia+H(z)+GRB) for the THDE with $Q=0$, and 1.581 (SNe Ia), 1.652 (SNe Ia+H(z)), and 2.214 (SNe Ia+H(z)+GRB) for the THDE with nonlinear interaction between dark sectors. It is interesting to mention that the highest deviation with respect to the $\Lambda CDM$ model at the present epoch relates to
the nonlinear interacting THDE with Ricci scale as L with $J_{0}$=2.214 for SNe Ia+H(z)+GRB data, whereas the lowest deviation corresponds to the $J_{0}$=1.574 with SNe Ia sample for the THDE with Ricci horizon (with $Q=0$).\\\\

\subsection{The $OM$ diagnostic}\label{subsec2}

Here, we benefit from the $OM$ as a geometrical diagnostic to display that the noninteracting THDE and nonlinear interacting THDE models will behave as the cosmological constant, quintessence, or phantom model in different redshifts. For this, if the evolution of the $OM$ does not show any slope, the models are the $\Lambda CDM$. But deviation from constant $OM$, namely positive slope or negative slope corresponds to the phantom or quintessence models. The relation of $OM$ diagnostic is given by (\cite*{bib60})

\begin{eqnarray}\nonumber
OM= \frac{(\frac{H(z)}{H_{0}})^{2}-1}{(z+1)^{3}-1}.
\end{eqnarray}

Where $H_{0}$ is the present value of the Hubble parameter and in our model using the best fit values of $h$ in Tables 2 and 3, it will be equal to $100\times h$.

The results of the $OM$ diagnostic can be observed in Fig. 9. The solid "GreenYellow" line in this figure belongs to the $\Lambda CDM$ model for the SNe Ia, SNe Ia+H, and the SNe Ia+H+GRB data (see Table 6).

In the upper panel related to the SNe Ia data for $OM$ plot, the THDE with noninteraction term (written as the THDE in the panel) behaves as the quintessence at $0.540< z\leq4.5$, and then, it behaves as the phantom at $-1\leq z<0.540$. This panel also shows the quintessential behavior of the nonlinear interacting THDE with Ricci cutoff (written as the NITHDE in the panel) at $0.523< z\leq4.5$ and phantom behavior at $-1\leq z<0.523$.

However, the middle plot belongs to the evolution of the $OM$ parameter with respect to the $z$ for our models (THDE and NITHDE models) with the SNe Ia+H sample. From this plot, the THDE with Ricci horizon experience the quintessence phase in $0.577< z\leq4.5$ and the phantom phase at $-1\leq z<0.577$. But, the nonlinear interacting THDE with Ricci as IR L shows the quintessence region for the model at $0.572< z\leq4.5$ and phantom regime at $-1\leq z<0.572$.

Finally, the lower graph of Fig. 9 describes the $OM$-$z$ plot in our models for the SNe Ia+H+GRB dataset. From this plot, the THDE with Ricci cutoff ($Q=0$) lies in the quintessence phase at $0.591< z\leq4.5$ and phantom state at $-1\leq z<0.591$. But, the nonlinear interacting THDE with Ricci scale as IR will behave as the quintessence at $0.582< z\leq4.5$ and phantom at $-1\leq z<0.582$.\\\\\\

\begin{table*}[t]
\caption{The best-fit parameters of $\Lambda CDM$ model }
\begin{tabular}{@{}lllllll|}
\hline
~Parameters&\hspace{0.9 cm}SNe Ia&\hspace{0.9 cm}SNe Ia+H(z)&\hspace{0.9 cm}SNe Ia+H(z)+GRB\\
\hline
\hline
~$h$~&\hspace{0.9 cm}0.6979&\hspace{0.9 cm}0.6974&\hspace{0.9 cm}0.6954\\
~$\Omega_{DE}$~ &\hspace{0.9 cm}0.707&\hspace{0.9 cm}0.703&\hspace{0.9
cm}0.684\\
~$\chi^{2}_{min}$~&\hspace{0.9 cm}1067.345&\hspace{0.9 cm}1078.997&\hspace{0.9 cm}1309.655\\
\hline
\end{tabular}
\end{table*}

\begin{figure*}
\centering
\includegraphics[width=0.55\textwidth]{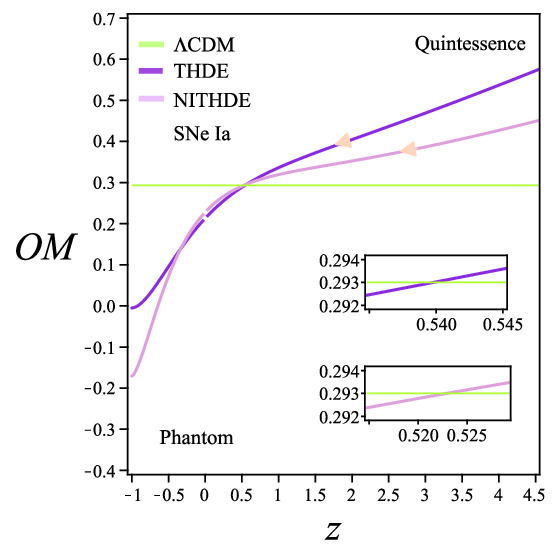}\\
\includegraphics[width=0.55\textwidth]{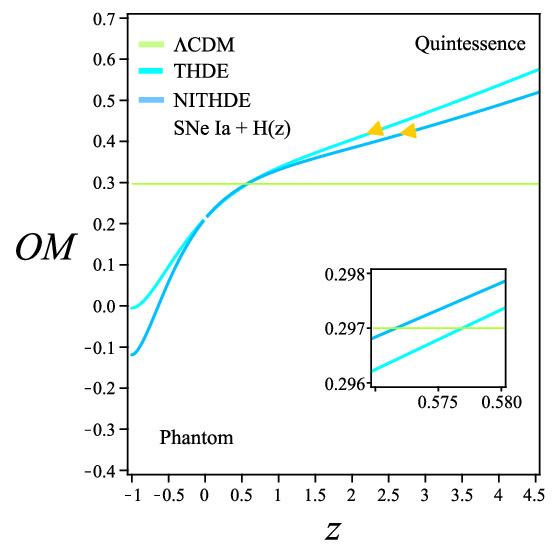}\\
\includegraphics[width=0.55\textwidth]{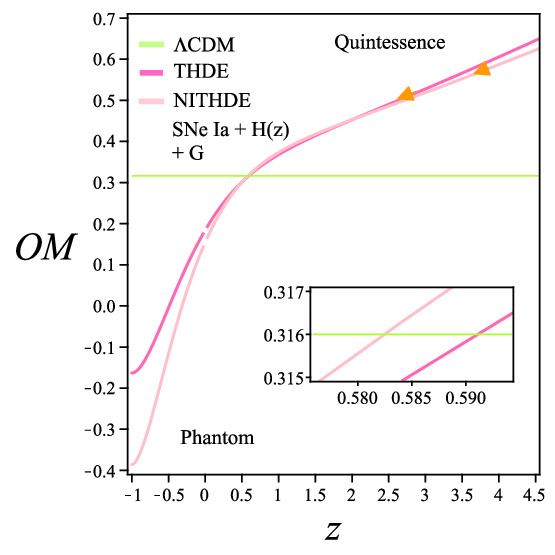}
 \caption{The evolution of $OM$ parameter versus redshift for the THDE without interaction term with
  Ricci cutoff (THDE) and the nonlinear interacting THDE with Ricci horizon as IR cut off (NITHDE)}
 \label{fig:9}       
\end{figure*}

\section{The age of the Universe}\label{sec7}

Now, we estimate the Universe age for the noninteracting and (nonlinear) interacting THDE with Ricci horizon utilizing the best fit values in Section 3. To measure The age of the Universe we use the following relation (\cite{bib22,bib61})

\begin{eqnarray}\nonumber
&&t= \int \frac{dt dH}{dH}\approx(\frac{1}{\frac{\frac{dH}{dX}}{H}})_{(z=0)}\int \frac{dH}{H^{2}}\approx\nonumber\\
&& \frac{1}{(2-\frac{(3 H_{0}^{2}\Omega_{DE0})^\frac{1}{2-\delta}}{H_{0}^{2} \lambda^\frac{1}{2-\delta}})}\frac{1}{H_{0}}
\end{eqnarray}

The results of the age of the Universe for the noninteracting THDE with Ricci as IR are obtained as $t\approx\frac{3.1}{H_{0}}$, $\frac{3.1}{H_{0}}$, and $\frac{3.6}{H_{0}}$ for the SNe Ia, SNe Ia + H(z), and SNe Ia + H(z) + GRB, respectively. As well as this, we have $t\approx\frac{2.9}{H_{0}}$, $\frac{3.1}{H_{0}}$, and $\frac{4.3}{H_{0}}$ for the SNe Ia, SNe Ia + H(z), and SNe Ia + H(z) + GRB samples for the nonlinear interacting THDE with Ricci cutoff. These results are comparable with the work done by \cite{bib61} for the THDE with Hubble cutoff with $t\approx\frac{2.74}{H_{0}}$ calculated using assumed values $\delta=2.1$, $\alpha=0.5$, $\beta=0.002$, and $t\approx\frac{3.28}{H_{0}}$ extracted from the fixed $\delta=2.1$, $\alpha=0.7$, $\beta=0.1$.

Although our results of the cosmic age are not close to observational results (such as Planck 2018 $TT+Low E$ with 13.825 Gyr), our models satisfy this principle in cosmology that the models should provide the Universe age higher than the age of Universe constituents (\cite{bib62,bib63,bib64}), which leads to avoiding the age problem too.

\section{Conclusions}\label{sec7}

In this paper, we have studied the observational limits on the THDE and nonlinear THDE with the Ricci cutoff as IR L using the SNe Ia, SNe Ia+H(z), and SNe Ia+H(z)+GRB datasets. We have graphed the 1$\sigma$ and 2$\sigma$ confidence levels and one-dimensional likelihood distribution.

At the continuation of our work, we have depicted the $H(z)-z$ plot in which the theoretical function of $H(z)$ has been compared with 27 measurements of the Hubble parameter $H(z)$. This graph identifies the suitable agreement between our models and the observational data. Apart from this, the $H(z)$ graph has demonstrated the turning point feature for our models in the future epoch.

After the $H(z)$ plot, we have analyzed the $\omega_{DE}$, $\omega_{eff}$, and $q$ parameters against redshift for the models. Investigating the evolution of $\omega_{DE}$ versus $z$ elucidates a quintessence-like behavior of dark energy in the THDE and (nonlinear) interacting THDE with the Ricci horizon at an early epoch and phantom-like behavior in the current era. After that, dark energy moves toward the $\Lambda$ barrier state in the future era.
 The analyzing of the $\omega_{eff}$ shows that this parameter passes the phantom divide line in the future and then tends to -1 at $z\longrightarrow-1$.

The classical stability trajectories with respect to redshift indicate that the noninteracting and nonlinear interacting THDE with Ricci cutoff behaved stably in the early time and unstably in the present and future eras.

In the next step of our work, to distinguish the THDE and nonlinear interacting THDE models with the $\Lambda CDM$ model, we have used the $J$ and $OM$ parameters. Both parameters denote that our models deviate from the $\Lambda CDM$ model during the Universe evolution. The $J$ parameter indicates that the noninteracting and nonlinear interacting THDE models (with Ricci scale as L) will move toward the $\Lambda CDM$ model in the late time. However, the $OM$ parameter shows more details rather than the $J$ parameter. From the $OM$ diagnostic tool, we have found that our models had a quintessence nature in the past, while it behaves as the phantom in the future era.

In the final step, we calculated the age of the Universe for our models. The results represent that our models lead to a cosmic age older than the age measured by observations. However, our models will not show the age problem.

\section*{acknowledgments}
I am incredibly grateful to the dear reviewer for the significant and valuable comments which caused this manuscript to improve considerably.

\section*{Declarations}

\begin{itemize}
\item Funding: This research received no funding.
\item Conflict of interest: The author declares that she has no conflict of interest.
\item Ethics approval: Not applicable
\item Consent to participate:  Not applicable
\item Consent for publication:  Not applicable
\item Availability of data and materials: This manuscript has no associated data or the data will not be
    deposited.
\item Code availability: Not applicable
\item Authors' contributions: This work completely has done by Zahra Feizi Mangoudehi.
\end{itemize}

\end{document}